\documentclass[a4paper,11pt]{article}
\usepackage{jheppub}

\usepackage{dsfont}
\usepackage{extarrows}
\usepackage[all]{hypcap}

\usepackage[T1]{fontenc} 
\usepackage[utf8]{inputenc}
\usepackage{hyperref}
\usepackage{placeins}
\usepackage{amssymb}
\usepackage[english]{babel}

\usepackage{latexsym}
\usepackage{amsmath}
\usepackage{amsfonts}
\usepackage{amssymb}
\usepackage{stmaryrd}
\usepackage{graphicx}
\usepackage{mathtools}
\usepackage{amsmath}
\usepackage{fontenc}
\usepackage{hyperref}

\usepackage{empheq}
\usepackage{paralist}

\usepackage{amssymb}
\usepackage{color}

\usepackage{pgf}
\usepackage{amscd}

\makeatletter
\renewcommand*\env@matrix[1][*\c@MaxMatrixCols c]{%
  \hskip -\arraycolsep
  \let\@ifnextchar\new@ifnextchar
  \array{#1}}
\makeatother

\def\be{\begin{equation}}
\def\ee{\end{equation}}
\def\ba{\begin{eqnarray}}
\def\ea{\end{eqnarray}}

\def\r{\rho}
\def\a{\alpha}

\def\b{\beta}

\def\G{\Gamma}

\def\c{\chi}

\def\m{\mu}
\def\n{\nu}

\def\L{\Lambda}

\def\G{ {\cal G}}

\def\R {\mathbb{R}}

\newcommand{\nc}{\newcommand}
\nc{\lb}{\llbracket}
\nc{\rb}{\rrbracket}
\nc{\gl}{\llbracket}
\nc{\gr}{\rrbracket}

\newcommand{\eq}[1]{\begin{equation}
                     \begin{split} #1 \end{split}
                     \end{equation}}

\allowdisplaybreaks[2]
\numberwithin{equation}{section}

\title{U-duality and Courant Algebroid in Exceptional Field Theory}

\author{Rui Sun}

\affiliation{Korea Institute for Advanced Study,\\
		85 Hoegiro, Dongdaemun-Gu, Seoul 02455, Korea}

\emailAdd{sunrui@kias.re.kr}

\abstract{
In this paper, we study the field transformation under U-duality in exceptional field theories. Take $SL(5)$ and $SO(5,5)$ exceptional field theories as examples, we explicitly show that the U-duality transformation is governed by the differential geometry of a corresponding Courant algebroid structure. The field redefinition specified by $SL(5)$ and $SO(5,5)$ transformations can be realized by Courant algebroid anchor mapping. Based on the existence of Courant algebroid in $E_d$ exceptional field theory, we expect that the Courant algebroid anchor mapping also exist in exceptional field theories with higher dimensional exceptional groups, such as $E_6$ and $E_7$. Intriguingly, the U-dual M2-brane and M5-brane can be realized with the same structure of Courant algebroid in exceptional field theory. Since in each exceptional field theory, all the involved fields can be mapped with the same anchor, the full Lagrangian is governed by the Courant algebroid anchor mapping. In particular, this is realized by the U-duality mapping in extended generalised geometry, from the extended bundle $E=TM \oplus \Lambda^2 T^*M \oplus \Lambda^5  T^*M\oplus \Lambda^6 TM$ to the U-dual bundle  $E^*=T^*M \oplus \Lambda^2 TM \oplus \Lambda^5 TM \oplus \Lambda^6 T^*M$ under the global  $E_d$ symmetry. From M-theory point of view, a U-dual effective theory of  M-theory is expected from Courant algebroid anchor mapping in such a global manner via U-duality. 
}

\begin{document}\maketitle

\section{Introduction}

In double field theory, the metric and Kalb-Ramond field are unified in a generalised metric. T-duality is manifested as global symmetry of the double field action, which incorporates both the type II supergravity and the T-dual $\beta$-supergravity~\cite{Hull:2004in,Dabholkar:2005ve,Grana:2008yw,Hull:2009mi,Hohm:2010jy,Hohm:2010pp,Geissbuhler:2013uka}.
The background possesses $d$ commuting isometries and the T-duality group is described by an $O(d, d)$ global symmetry. Such that the double field theory spacetime become a doubled spacetime with $O(d, d)$ global symmetry manifested. The local  $O(d)\times O(d)$ symmetry is equivalent to the Lorentz group for the doubled space. 
T-duality as a perturbative symmetry in string theory is represented by the global symmetry $O(d, d)$, while to include non-perturbative symmetry S-duality in one theory, this calls for a more general new symmetry group. 
M-theory unifies T-duality and S-duality into U-duality in such a manner. The ``winding and momentum modes'' under U-duality involves not only the strings but also the winding of branes of different world-volume dimensions, such as M2-brane and M5-brane. The brane winding plays an important role and the resulting symmetry group becomes dimensionally dependent. 
The U-duality group $G$ is associated with an reduction on a $d$ dimensional torus, and given by exceptional group $E_{d}$ with $d$ as the internal dimension~\cite{Cremmer:1979up, deWit:1985iy, deWit:1986mz, Berman:2013eva}. The maximal compact subgroup of  $E_{d}$ is the local Lorentz group $H$, together with the diffeomorphisms and $p$-form gauge transformations. The resulting scalar fields from the reduction inhabit the $G/H$ coset. 

Generalised geometry was constructed with a generalised bundle  $TM\oplus T^*M$~\cite{Hitchin:2003cxu, Gualtieri:2004}. 
This provides a structure on a $d$ dimensional manifold $M$ with a natural action of the group $O(d,d)$.
The $O(d,d)$ global symmetry transforms between the type II supergravity and the dual non-geometric $\beta$ supergravity, and maps between the tangent and cotangent bundles in the generalised geometry. 
One considers a $d$ dimensional manifold $M$ equipped with generalised tangent bundle  $E={TM}\oplus {T^*M}$ which corresponding to the T-duality in double field theory~\cite{Grana:2008yw}. 
The T-dual framework realized by the global $O(d, d)$ transformation are known as the so-called non-geometric framework.  
An intriguing structure reveals that the $O(d, d)$ induced field redefinition can be associated to a corresponding Lie algebroid and the redefined supergravity action is governed by the differential geometry of the Lie algebroid~\cite{Blumenhagen:2013aia}. This structure was further generalised to heterotic double field theory with abelian gauge field incorporated in~\cite{Blumenhagen:2014iua}. This generalization is  straightforward by extending the global
symmetry group from $O(d, d)$ to $O(d, d+n)$ with $n$ gauge fields.  Heterotic T-dual of a constant gauge flux background was also given therein. For every gauge field $A^\alpha$ a new coordinate $y^\alpha$  is introduced, and the generalised metric is also generalised to include the gauge fields~\cite{Siegel:1993xq, Siegel:1993th, ReidEdwards:2008rd, Hohm:2011ex}.

The M-theory reduced lower dimensional supergravity with $E_d$ symmetry is known as exceptional field theory as it calls for an  generalised geometry with exceptional algebra structure~\cite{Hull:2007zu, PiresPacheco:2008qik}. The associated algebroid turns out to be a Courant algebroid for the extended bundle $E=TM \oplus \Lambda^2 T^*M \oplus \Lambda^5  T^*M\oplus \Lambda^6  TM$ instead of Lie algebroid for tangent bundle $TM$~\cite{Courant:90, Gualtieri:2003dx, Gualtieri:2004, Berman:2011cg}.
Similar as in double field theory, the bosonic sector of the kinetic term can be entirely realized by  exceptional field theory in terms of a generalised metric on an extended space~\cite{Berman:2010is, Berman:2011pe, Thompson:2011uw}.
The local gauge symmetries not just incorporates the 2-form gauge symmetries of string theory but simultaneously involve the 3-form and 6-form gauge symmetries of eleven dimensional supergravity. A generalised Lie derivative generates these local transformation entirely. The closure of gauge algebra imposes a closure constraint on exceptional field theory realizing U-duality of M-theory.

By considering the eleven dimensional supergravity compactifies on $d$ dimensional torus $T^d$, the U-duality in M-theory  is manifested with an exceptional symmetry group $E_d$~\cite{Cremmer:1979up, deWit:1985iy, deWit:1986mz, Berman:2013eva}. The dimension reduction of supergravity results in a hidden $E_{d}$ global symmetry.
This provides a unified description of bosonic eleven dimensional supergravity with the restriction to a $d$ dimensional manifold for all $d \leq 7$~\cite{Coimbra:2011ky}. These exceptional field theories are  based on extended spaces which admit  natural $E_{d}\times R^+$ actions. Similar to double field theory, the bosonic degrees of freedom can be parametrized into a ``generalised metric'', while the local $O(d)$ symmetry is generalised to $H_d$, the maximally compact subgroup of $E_{d}$.   
This $E_d$ group acts on the metric $g$ and the M2-brane and M5-brane~\cite{Cremmer:1979up, Julia:sugra}. 
Here $E_d$ represents the exceptional group with $d=2,...,7$, and  their maximal compact subgroups $H_d$ discussed in~\cite{Keurentjes:2003hc},\cite{Keurentjes:2003yu}, together with the cosets $E_d/H_d$ moduli are given in Table~\ref{tab1}.  These exceptional groups are found to be the symmetries of reduced eleven dimensional supergravity theories~\cite{Cremmer:1979up, deWit:1985iy, deWit:1986mz}. Generalized Cartan Calculus for exceptional duality groups were constructed in~\cite{Hohm:2015xna,Wang:2015hca}.
The generalised tangent bundle $E$ is constructed over the  $d$ dimensional manifold ${\cal M}$ with fibre in the representation ${\bf R}$ of $E_{d}$ . 
 \begin{table}[] \label{tab1} 
	\begin{center}
		\begin{tabular}{|c|cccc|}
			\hline
			d & $E_{d}$ & $H_d$ & $E$ & $E_d/H_d$ moduli\\
			\hline
			2  & $SL(2,\R) \times \R$ & $SO(2)$ & $ T\oplus \L^2T^*  $ & $G$\\
			3  & $SL(3,\R) \times SL(2,\R)$ & $SO(3) \times SO(2)$  &$ T\oplus \L^2T^*  $ & $G, C_3$\\
			4  & $SL(5,\R)$  & $SO(5)$ & $ T\oplus \L^2T^*   $ & $G,C_3$ \\
			5  & $SO(5,5)$ & $SO(5) \times SO(5)$ & $ T\oplus \L^2T^*  \oplus \L^5T^* $ & $G,C_3$\\
			6  & $E_{6}$  & $USp(8)$ & $ T\oplus \L^2T^*  \oplus \L^5T^* \oplus \L^6T $  & $G,C_3,C_6$\\
			7  & $E_{7}$ & $SU(8)$  & $  T\oplus \L^2T^*  \oplus \L^5T^* \oplus \L^6T $  & $G,C_3,C_6$\\
			\hline    
		\end{tabular}
		\caption{The U-duality groups $E_d$, their maximal compact subgroups $H_d$, the cosets $E_d/H_d$ moduli, and the bundle $E$ over the  $d$ dimensional manifold ${\cal M}$ with fibre in the representation ${\bf R}$ of $E_{d}$.
		}
	\end{center}
\end{table}

For M-theory, generalised geometry introduced by Hitchin was further extended to manifest the symmetries of U-duality in an extended version~\cite{Hull:2007zu, PiresPacheco:2008qik, Berman:2010is}. The eleven dimensional supergravity fields are constructed in generalised tangent space, {\emph e.g.}, $ E \simeq TM \oplus \Lambda^2T^*M \oplus \Lambda^5T^*M \oplus \Lambda^6 TM $, admitting the $E_{d}$ structure for eleven dimensional supergravity restricted to $d \leq 7$ dimensions~\cite{Coimbra:2011ky}. We consider the dual generalised tangent space $E^* \simeq T^*M \oplus \Lambda^2 TM \oplus \Lambda^5 TM \oplus \Lambda^6 T^*M$.
This generalised the original version of generalised tangent space $E \simeq TM \oplus T^*M$ introduced by Hitchin~\cite{Hitchin:2003} which was utilized in double field theory. In exceptional field theory, the world-sheet doubled formalism describes strings and M-branes in an extended spacetime with extra coordinates.

Consider eleven dimensional supergravity compactifies on a $d$ dimensional torus  $T^d$, the global duality group(representing the U-duality group) of the maximal supergravity is the maximal real subgroup of $E_d$.
When we have four and five dimension reduction, the resulting theories are seven and six dimensional maximal supergravity with $SL(5)$ and $SO(5,5)$ bosonic sectors. The so-called ``generalised metric'' incorporated metric $g$, and higher forms represents the M-branes. The fields transform under $SL(5)$ and $SO(5,5)$ group transformations.

In this paper, we point to the question that whether the U-duality in M-theory, in the present of $E_d$ transformation in exceptional field theory, underline an algebroid structure as in double field theory~\cite{Blumenhagen:2013aia, Blumenhagen:2014iua}. Moreover, we study the field redefinition under U-duality via Courant algebroid anchor mapping.
We will in detail study the $E_d$ group transformation in $E_4=SL(5)$ and $E_5=SO(5, 5)$ exceptional field theories. In Section~\ref{sec:sl5}, we perform the $SL(5)$ group transformation to the generalised metric of $SL(5)$ exceptional field theory, and give the field redefinitions in the U-dual frame under Courant algebroid structure. In Section~\ref{sec:so55}, we study the  $SO(5,5)$ global transformation to the generalised metric of  $SO(5,5)$ exceptional field theory, and derive the field redefinition in the algebroid anchor manner. In Section~\ref{sec:algebroid}, we show that the Courant algebroid indeed govern the anchor structure we obtain in $SL(5)$ and $SO(5,5)$ exceptional field theories. In Section~\ref{sec:e67}, we show that  $E_6$ and $E_7$ exceptional field theories are expected to be associated with Courant algebroid as well. The action of dual exceptional field theory is also expected to be governed by the Courant algebroid structure.

\section{Lie Algebroid and Courant Algebroid}
\label{sec:algebroid}

In double field theory, the field redefinition under T-duality\,(represented by global symmetry) is governed by a Lie algebroid structure, which admits a natural generalization of the usual differential geometry. Thus, the covariant derivatives, torsion and curvature tensors can also be defined under a Lie algebroid anchor mapping~\cite{Halmagyi:2008dr, Halmagyi:2009te,  Berman:2010is,  Blumenhagen:2012pc, Blumenhagen:2013aia, Blumenhagen:2014iua}.

Before studying the corresponding algebroid structure in exceptional field theory, we first review the Lie algebroid structure in double field theory. T-duality is manifested as global $O(d, d)$ symmetry, while the transformation matrix takes the general form of 
\eq{
	\label{abcd}
	U =\begin{pmatrix} 
		a^a{}_{b}  &~b^{a\beta} 
		\\c_{\alpha b} &~d_{\alpha}{}^{\beta}
	\end{pmatrix} \,.
}
The $O(D,D)$ transformation matrix \eqref{abcd} acts on a tuple $(X^a,\xi_{\alpha})^t$, 
with $X=X^a e_a$ a vector field and $\xi=\xi_{\alpha} e^{\alpha}$ a one-form.
The submatrices $a^a{}_{b}, b^{a\beta}, c_{\alpha b}, d_{\alpha}{}^{\beta}$ denote the linear mappings 
\eq{	
	\arraycolsep2pt
	\begin{array}{r@{\hspace{6pt}}lcl@{\hspace{50pt}}r@{\hspace{6pt}}lcl}
		a: & TM &\to& TM\,, &
		b: & T^*M &\to& TM \,, 
		\\[2mm]
		c: & TM &\to& T^*M \,, &
		d: & T^*M & \to & T^*M \,.
	\end{array}
}
An anchor map $\rho$ can be utilized to map the elements of the Lie algebroid bundle $E=TM$ to the tangent bundle $TM$. The inverse anchor $\rho^{-1}$ maps the tangent bundle $TM$ to the Lie algebroid bundle $E=TM$. A bracket for the Lie algebroid bundle can be determined following the original Lie bracket~\cite{Blumenhagen:2013aia}. Such relation between different brackets is established by the anchor $\rho$ as well.  

In particular, with the index structure of anchor as 
\eq{
	\rho \equiv \rho_\alpha{}^a\,, \hspace{20pt}
	\rho^{-1}\equiv (\rho^{-1})_a{}^\alpha\,, \hspace{20pt}
	\rho^t \equiv (\rho^t)^a{}_\alpha\,,\hspace{20pt}
	\rho^*\equiv (\rho^*)^\alpha{}_a \,,
}
the index $\alpha$ corresponds to the Lie algebroid,  
while the index $a$ gives the index in $TM$. The homomorphism mapping in between the $TM$ and $T^*M$ can be denoted by
\eq{	
	\arraycolsep2pt
	\begin{array}{r@{\hspace{6pt}}lcl@{\hspace{50pt}}r@{\hspace{6pt}}lcl}
		\rho_a{}^b : & TM &\to& TM\,, &
		\rho^{}{}_{a\alpha}: & T^*M &\to& TM \,, 
		\\[2mm]
		\rho^{\alpha a}{}_{}: & TM &\to& T^*M \,, &
		\rho^\alpha{}_\beta: & T^*M & \to & T^*M \,.
	\end{array}
}
On the left-hand side we have the standard tangent and cotangent bundles of the manifold,  while the right-hand side gives the Lie algebroid bundles $TM$ and $T^*M$.

\begin{figure}[h]
	\centering
	\includegraphics[width=0.9\textwidth]{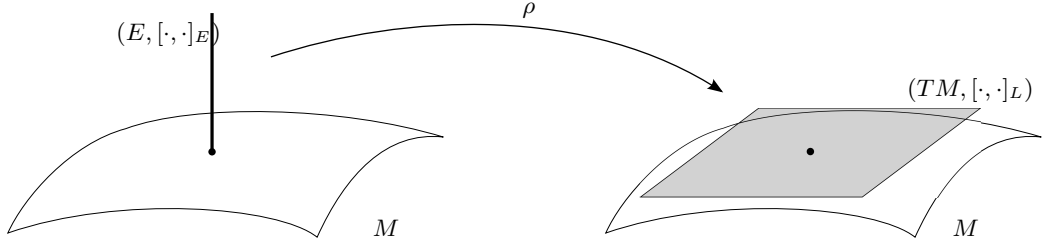}
	\begin{picture}(0,0)
		\put(-198,85){\footnotesize $\rho$}
		\put(-254,1){\footnotesize $M$}
		\put(-37,1){\footnotesize $M$}
		\put(-54 ,54){\footnotesize $(TM,[\cdot,\cdot]_L)$}
		\put(-350,75){\footnotesize $(E,[\cdot,\cdot]_E)$}
	\end{picture}
	\vskip5pt
	\caption{Lie algebroid illustrated by Blumenhagen, Deser, Plauschinn, and  Rennecke~\cite{Blumenhagen:2013aia}. \label{fig_01}}
\end{figure}

The Lie algebroid can be considered as a generalization of a Lie algebra for which the structure constants are space-time dependent.
Moreover, it is characterized by a vector bundle $E$ over a manifold $M$,  a bracket $[\cdot,\cdot ]_E : E \times E \rightarrow E$, as well as a homomorphism $\rho : E \rightarrow TM$ denoted by the anchor.
As illustrated in figure~\ref{fig_01}, on the left side, a manifold $M$ is illustrated with a bundle $E$, together with a bracket $[\cdot,\cdot]_E$ mapped via the anchor $\rho$ to the right side tangent bundle $TM$ with Lie bracket $[\cdot,\cdot]_L$.

Similar to the Lie bracket,  the bracket $[\cdot,\cdot]_E$ is required to satisfy a Leibniz rule.
Defining $f\in {\cal C}^{\infty}(M)$ as  the functions and $s_i$ for the sections of $E$, we obtain the Leibniz rule represented by
\eq{
	\label{Leibniz}
	[ s_1, f s_2]_E = f \hspace{1pt}[ s_1,s_2]_E + \rho(s_1)(f) s_2  \,,
} 
where $\rho(s_1)$ is a vector field which acts on $f$ as a derivation.
If  the bracket $[ \cdot,\cdot]_E$  in addition admits a Jacobi identity
\eq{
	\label{jacobii}
	\bigl[  s_1, [ s_2, s_3 ]_E \bigr]_E = \bigl[ [  s_1, s_2 ]_E , s_3 \bigr]_E +  \bigl[ s_2, [ s_1, s_3 ]_E\bigr]_E\, ,
}
then $(E,[\cdot,\cdot]_E, \rho)$ is  called
a \emph{Lie algebroid}. Note that if the Jacobi identity is not satisfied, the resulting structure becomes a \emph{quasi-Lie algebroid} instead.

Furthermore, the Lie algebroid vector fields and their Lie bracket $[\cdot,\cdot]_L$ can be represented by sections of $E$ and the corresponding bracket $[\cdot,\cdot]_E$. Meanwhile, the relation between these different brackets and the sections is defined by the anchor mapping. Then requiring that the anchor $\rho$ is a homomorphism implies that
\eq{	\label{anhom}
	\rho\bigl( [ s_1,s_2]_E\bigr) = \bigl[\rho(s_1),\rho(s_2)\bigr]_L \,. }
By considering the tangent bundle $E=TM$ with the usual Lie bracket $[ \cdot,\cdot]_E=[\cdot,\cdot]_L$ and trivial anchor being the identity map, i.e.,\ $\rho={\rm id}$,
under conditions \eqref{Leibniz}, \eqref{jacobii}, and \eqref{anhom}, $E=(TM,[\cdot,\cdot]_{L},\rho= \textrm{id})$ is indeed a Lie algebroid~\cite{Blumenhagen:2013aia}, .

For cotangent bundle $E=T^*M$, one considers a Poisson manifold $(M,\beta)$ with Poisson tensor $\beta =  \frac{1}{2}\,\beta^{ab}e_a \wedge e_b$, in which $\{e_a\}$ denotes the basis of vector fields.
A Lie algebroid is constructed by defining
$E=(T^* M,[\cdot,\cdot ]_K,\rho =  \beta^\sharp)$, in which
the anchor $\beta^{\sharp}$ is given by
\eq{ 
	\beta^{\sharp} (e^a) = \beta^{am}e_m \,,
}
in which $\{e^a\}\in\Gamma(T^*M)$ being the one-form basis dual to the basis of the vector fields.
Here the \emph{Koszul bracket} on $T^*M$ is denoted by $[\cdot,\cdot ]_K$. For one-forms $\xi$ and $\eta$ this yields to
\eq{
	\label{koszul}
	[\xi,\eta]_K = L_{\beta^\sharp(\xi)}\eta
	-\iota_{\beta^\sharp(\eta)}\,d\xi \, ,
}
where the Lie derivative on forms is defined by $L_{X} = \iota_X \circ d + d \circ \iota_X$ with $d$ as the de Rham differential. 
For  one forms $\xi=\xi_a dx^a$ and $\eta=\eta_a dx^b$, with basis $\{dx^a\}$, the Koszul bracket is then defined by $[\xi,\eta]_K =\left(\xi_a \beta^{ab} \partial_b \eta_m - \eta_a \beta^{ab} \partial_b \xi_m + \xi_a \eta_b \partial_m \beta^{ab}\right) dx^m$.
Recall that $\beta$ is a Poisson tensor, \emph{i.e.}, $\beta^{[a|m}\partial_m \beta^{|bc]}=0$, one can verify that the conditions \eqref{Leibniz}, \eqref{jacobii} and \eqref{anhom} are satisfied. Thus a Lie algebroid does exist for cotangent bundle as well.

Now the question becomes whether such an algebroid structure also exists and governs the U-duality transformation in exceptional field theory. This will be studied below.

\subsection*{Courant Algebroid and Bracket}

Recall the fact that when there is a mapping $f$ for vector space $V$ and $W$, there exist
\eq{	
	\arraycolsep2pt
	\begin{array}{r@{\hspace{6pt}}lcl@{\hspace{50pt}}r@{\hspace{6pt}}lcl}
		&	(f: V \to \mathbb{R}) \Leftrightarrow f \in V^*,\, e.g., V=TM, 
		\\[2mm]
		&	(f: V \to \mathbb{W}) \Leftrightarrow f \in V^* \otimes W,
		\\[2mm]
		& g: V\otimes V \to \mathbb{R} \Leftrightarrow g \in V^*\otimes V^*,
	\end{array}
} with metric represented by $g$ mapping from $V\otimes V$ to $\mathbb{R}$.
For the homomorphism mapping $\gamma$ involved in our discussion, we have 
\eq{	
	\arraycolsep2pt
	\begin{array}{r@{\hspace{6pt}}lcl@{\hspace{50pt}}r@{\hspace{6pt}}lcl}
		&	\gamma: TM \to T^* M \Leftrightarrow \gamma \in T^*M\otimes T^* M,
		\\[2mm]
		&	g: TM \otimes TM \to \mathbb{R} \Leftrightarrow g \in T^*M \otimes T^* M.
	\end{array}
}

In exceptional field theory, the generalised geometry is extended with higher forms, which appear as section of $\Lambda^2 T^*M$, $\Lambda^5 T^*M$, and so on~\cite{Hull:2007zu}. 
It is obvious that when the off-diagonal components in the transformation matrix of $E_d$ are non-trivial,  the mapping denoted by off-diagonal element $b$ and $c$ in the transformation matrix represent as, \emph{e.g.},
\eq{\label{TMdelta}	
	\arraycolsep2pt
	\begin{array}{r@{\hspace{6pt}}lcl@{\hspace{50pt}}r@{\hspace{6pt}}lcl}
		&	b: \Lambda ^2 T^* M \to TM \Leftrightarrow b\in \Lambda ^2 T M \otimes TM,
		\\[2mm]
		&	c: TM \to \Lambda ^2 T^* M  \Leftrightarrow c \in T^*M \otimes \Lambda ^2 T^* M ,
	\end{array}
}
where $b$ and $c$ represent the mapping in between  $TM$ and $\Lambda^2 T^*M$. 
The extended tangent bundle includes $TM\oplus\Lambda^2 T^*M$ with $\Lambda^2 T^*M$ corresponding to a M2-brane charge. 
Under the mapping $TM \to E=TM\oplus\Lambda^2 T^*M$,  the metric $g$ on $TM$  is mapped to  the metric on $E = TM \oplus \Lambda^2 T^*M$ over the manifold $M$. The pullback of the metric on $TM$ obtain an extra contribution from  $\Lambda^2 T^*M$.
However, this mapping  is governed beyond the structure of Lie algebroid, yet with Courant algebroid utilized.
In Section~\ref{sec:sl5} and \ref{sec:so55}, we will explicitly show the Courant algebroid anchor mapping in $SL(5)$ and $SO(5,5)$ exceptional field theories.

Recall that in exceptional field theory, the algebroid structure is associated with extended generalised bundle with wedge products of $TM$ and $T^*M$, such as $\Lambda^2 T^*M, \Lambda^5 T^*M$~\cite{Hull:2007zu}. This necessarily calls for the corresponding generalised algebroid structure. Along this direction, Courant bracket was firstly introduced by Courant for $TM\oplus \Lambda^p T^*M$ with $p=1$~\cite{Courant:90}. For exceptional field theory, the Courant bracket is defined on pairs $V=(v, \mu)$ of a vector field $v$ and a $p$-form $\mu$ on a manifold $M$~\cite{Hitchin:2003cxu, Gualtieri:2003dx, Berman:2010is, Berman:2011cg, Arvanitakis:2018cyo, Arvanitakis:2019cxy}. 

The generalised Lie derivative in the extended generalised geometry are defined with reparametrisations of the generalised coordinates~\cite{Grana:2008yw}. The generalised vector fields are $V=(v, \mu)$ and $X=(x, \lambda),$ for which the $v$ and $x$ are vector fields and $\mu$ and $\lambda$ are $2$-forms. Therefore, the generalised Lie derivative is defined with the generalised vector $V$ as
\begin{equation}
	\hat{\mathcal{L}}_{X} V = \mathcal{L}_{x} v + \mathcal{L}_{x} \mu - i_{v} d \lambda.
	\label{defgenLie}
\end{equation}
This reproduces the transformation of a vector field under coordinate transformations. The second term on the right hand side of~\eqref{defgenLie} represents the transformation of a two-form field, $\mu,$ under coordinate transformations, while the third term represents the gauge transformations generated by $2$-forms $\lambda$.  

The corresponding Courant bracket for the extended generalised geometry is an operation defined with the antisymmetrisation of the generalised Lie derivative
\begin{align} 		\label{Courant}
	[X,Y]_{C} &= \frac{1}{2} \left(\hat{\mathcal{L}}_{X} Y - \hat{\mathcal{L}}_{Y} X\right) \\ \nonumber 
	&= [x,y] + \mathcal{L}_{x} \eta - \mathcal{L}_{y} \lambda +\frac{1}{2} d \left( i_{y} \lambda - i_{x} \eta \right), 
\end{align}
in which $X=(x,\lambda), \ \ Y=(y, \eta)$. The bracket on the right hand side, $[x,y]$, denotes the Lie bracket of vector fields, in which $x$ and $y$ represents the algebra of diffeomorphisms~\footnote{The bracket with lower index $C$ denotes the Courant bracket, while without the lower index it denotes the Lie bracket.}. Moreover, $\mathcal{L}_{x} \eta - \mathcal{L}_{y} \lambda$ is the gauge transformation followed by a diffeomorphism. The last term, $\frac{1}{2} d \left( i_{y} \lambda - i_{x} \eta  \right)$, turns out to be exact and thus has no impact on the $3$-form potential $C$. 
Therefore, the Courant bracket of such generalised diffeomorphisms\,(constructed with $TM\oplus \Lambda^2 T*M$) incorporates diffeomorphisms and gauge transformations. 

In order to provide a proper definition of Courant algebroid, we now check whether Jacobi identity is satisfied for exceptional field theory. 
The Jacobiator written in terms of the Courant bracket is given by
\begin{equation}
	J(X,Y,Z) = [[X,Y]_C,Z]_C + [[Y,Z]_C,X]_C + [[Z,X]_C,Y]_C,
\end{equation}
with $X=(x,\lambda), Y=(y,\mu)$ and $Z=(z,\kappa)$.
Though the Jacobi identity is not guaranteed in general, 
the Courant Jacobiator takes an exact form of
\begin{equation}\label{bid}
	J(X,Y,Z) = \frac{1}{4}d[(\iota_xL_y-\iota_yL_x)\kappa + (\iota_yL_z-\iota_zL_y)\lambda +(\iota_zL_x-\iota_xL_z)\mu],
\end{equation}
has been shown in~\cite{Berman:2011cg, Courant:90, Hitchin:2003cxu, Gualtieri:2003dx}. 

In exceptional field theory, for the fields that do not transform under gauge transformation, the Jacobi identity holds trivially.  For the field $C$, which is the only field changes under gauge transformation,  we have $\delta\,C = d\,\lambda$.   Since the Jacobiator takes an exact form, we have $\delta_{J(X,Y,Z)}\,C = 0$. 
Therefore, the Jacobi identity holds while acting on fields in exceptional field theory. And a Courant algebroid can be defined in exceptional field theory with the appearance of the 3-form potential $C$. 
In more detail, here we take $SL(5)$ exceptional field theory as an example. We use $X$, $Y$, $Z$ as $SL(5)$ bivectors with index structure $X^{ab}=X^{[ab]}$. The generalised vector field  $X=(x^i,\lambda_{ij})$, where $i, j = 1, ...4$ consequently. The $SL(5)$ indices on $X$ can be represented by 
\begin{equation} \label{genfield}
	X^{a b}=
	\begin{cases}
		X^{i5} = x^{i} , \\
		X^{5i} = -x^{i}, \\
		X^{ij} = \frac{1}{2} \eta^{ijkl} \mu_{kl},
	\end{cases}
\end{equation}
with $a,b =1,...5$.
It was proved in~\cite{Berman:2010is, Berman:2011cg} that the algebra closure of the generalised diffeomorphisms 
\eq{
	[\hat{\mathcal{L}}_{X},\hat{\mathcal{L}}_{Y}] V = \hat{\mathcal{L}}_{[X,Y]_C} V }
 requires the original section condition for $SL(5)$ exceptional field theory
\begin{equation}
	\partial_{[ab} \partial_{cd]} X =0 \qquad \text{and} \qquad  \partial_{[ab} X \cdot \partial_{cd]} Y  =0,
	\label{seccon}
\end{equation}
in which $X$ and $Y$ are arbitrary fields. This restriction was often utilized in a strong version  with $	\frac{\partial}{\partial y_{ij}} =0 $ for simplification. However, note that the indices $ab$ as given in~\eqref{genfield} are not simply the indices from the extended space with $X^{ij}, i,j =1,...4$~(corresponding to the bundle $\Lambda^2 T^*M$), but also  the ordinary indices $x^{i}, i=1,...4$ corresponding to the bundle $TM$.
The original section condition~\eqref{seccon}, which is derived according to the closure of the algebra, only requires that the fields with dependences on \emph{both} $TM$ and $\Lambda^2T^*M$ bundles  in the antisymmetric form vanish, not that the dependence on the extended dependences themselves vanish completely.
The solution for the original section condition~\eqref{seccon} in SL(5) and SL(N) theories were also discussed in~\cite{Blair:2013gqa,Park:2013gaj,Park:2014una}.
In double field theory, the original section condition corresponds to the closure constraint, or quadratic constraint as it is derived from the closure of the algebra. This closure constraint imposed on the gaugings 
was also shown to be weaker than the strong constraint\,(strong version of section condition) in the duality twist in double field theory~\cite{Hull:2009mi,Hohm:2010jy,Hohm:2010pp,Geissbuhler:2013uka}.

In exceptional field theory, although the original section condition~\eqref{seccon}\,(closure constraint) imposes restriction on the field transformations, 
it does allow the mapping between $E= TM \oplus \Lambda^n T^*M$ and $E^*= T^*M \oplus \Lambda^n TM$ according to the Courant algebra closure constraint.
Although the cross mapping between $\Lambda^mT^*M$ and $\Lambda^nT^*M$ with different $m$ and $n$ is forbidden from the closure of the algebra, the condition to perform the overall mapping between $E$ and $E^*$ is exactly the original section condition~\eqref{seccon}.  This is one of the key point we explore in this paper.   
	
From the global $E_d$ transformation point of view, we perform general  mapping and transformation, then present the allowed field redefinitions  as examples for $SL(5)$ and $SO(5,5)$ exceptional field theory.
Note that although we take particular choices of the transformation matrix, the Courant algebroid anchor can be differently chosen and thus result in different  field redefinitions in the ``non-geometric'' U-duality frame. In double field theory, such different anchor and different field redefinitions was explicitly shown with examples in~\cite{Blumenhagen:2013aia}.

\section{U-duality in M-theory}

In M-theory, the 3-form gauge potential $C_{abc}$ presents  in the  kinetic term of the effective action, represented by
\eq{\label{kin}
I_{kin} = -\int_{\cal M} d^dx\,\sqrt{g}\ \frac{1}{48}F_{abcd}F^{abcd}
}
with the field strength $F_{abcd}$ derived from the exterior derivative
of the 3-form abelian gauge field $C_{abc}$ as 
\eq{
F_{abcd} = 4\partial_{[a} C_{bcd]}.
}
Meanwhile, the kinetic term  $I_{kin}$ is gauge invariant under the gauge transformation,
\eq{
C_{abc} \rightarrow C_{abc} + 3\partial_{[a} \Lambda_{bc]}.
}
In Section~\ref{sec:algebroid}, we have presented that under such gauge transformation, the Courant algebroid holds for the field $C_{abc}$.

On the other hand, the Chern-Simons term is represented by
\eq{
I_{CS} = \lambda \int_{\cal M} d^dx \ \eta \ C_{abc}F_{defg}F_{hijk}
}
in which $\eta$ is the eleven-dimensional alternating tensor $\eta^{abcdefghijk}$ and  $\lambda$ is determined by supersymmetry. On the other hand, the Chern-Simon term is defined in eleven dimension only. Therefore, in lower dimensional exceptional field theory, we only consider the duality in kinetic terms in the following.

From the duality point of view, the 3-form $C_{abc}$ can be split into pure spatial components can the else, namely
\eq{
C_{abc} \to C_{ijk}, C_{0j}=B_{ij},
}
where the field strength of $C_{ijk}$
\eq{
F_{ijkl}= 4 \partial_{[i} C_{jkl]}
}
analogous to the magnetic field.
The field strength of $B_{ij}$ is analogous to the dual electric field,
\eq{
	G_{ijk}= 3 \partial_{[i} B_{jk]}.
}
Therefore, the kinetic term of the action can be decomposed into the metric and the 3-form fields~\cite{Duff:1990hn, Berman:2010is}. The M2-brane is constructed in the spacetime with the metric $g_{ab}$ and 3-form $C_{abc}$. 
For the bosonic part of the M2-brane, the action is given by
\eq{
	\int d^3\xi \sqrt{-h} \ 
	\bigl(\frac{1}{2}h^{\mu\nu}\partial_\mu X^a
	\partial_\nu X^b g_{ab}  +
	\frac{1}{6}\epsilon^{\mu\nu\rho}\partial_\mu X^a
	\partial_\nu X^b \partial_\rho X^c C_{abc}- \frac{1}{2}\bigr),
	\label{eq:M2action}
}
in which $\xi^\mu$ are world-volume co-ordinates and $h_{\mu\nu}$ is the  world-volume
metric with signature $(-++)$, and $\epsilon^{\mu \nu \rho}$ denotes the alternating tensor.
The electromagnetic duality exchange the roles of Bianchi identities and the equations of motion. In M-theory, this can be incorporated into matrix 
\eq{
\begin{pmatrix}
	G_{\mu a}\\ G_\mu^{mn}
\end{pmatrix} = \begin{pmatrix} g_{ab}+\frac{1}{2}C_a{}^{ef}C_{bef}&
	\frac{1}{\sqrt{2}}C_a{}^{kl}\\ \frac{1}{\sqrt{2}}C^{mn}{}_b & g^{mn,kl}
\end{pmatrix}\begin{pmatrix}F_\mu^b\\ F_{\mu kl}\end{pmatrix},  
\label{genmet}
} with $F_{\mu\ ab} = \partial_\mu y_{ab}$, and $ g^{mn,kl}=\frac{1}{2}(g^{mk}g^{nl}-g^{ml}g^{nk})$  raising an antisymmetric pair of indices. 
A generalised field strength can be defined as 
\eq{
	G_{\mu M} = \begin{pmatrix}G_{\mu a}\G_\mu^{ab}\end{pmatrix},
}
while the generalized displacement is represented by
\eq{
	F_\mu^M = \begin{pmatrix}F_\mu^a\ F_{\mu ab}\end{pmatrix}.
}
The equation of motion and the Bianchi identity can be unified into 
\eq{
\partial_\mu G^\mu_M = 0,
} and the field strengths can be manifested into
\eq{
G_{\mu M} = M_{MN} F^N_\mu,
\label{genmet1}
}
in which
\eq{\label{genm}
M_{MN} = \begin{pmatrix} g_{ab}+\frac{1}{2}C_a{}^{ef}C_{bef}&
		\frac{1}{\sqrt{2}}C_a{}^{kl}\\ \frac{1}{\sqrt{2}}C^{mn}{}_b & g^{mn,kl}
	\end{pmatrix}
}
 is called as the generalized metric incorporating the metric and 3-form gauge fields. 
The M-theory action can be expressed with the generalised field strength and generalised metric as
\eq{\label{lag}
	L = G^\mu_M F_\mu^M.
}
Consider M-theory compactifies on a $d$ dimensional torus $T^d$ with  $d \leq 7$, the eleven dimensional supergravity is reduced to a lower dimensional exceptional field theory with a hidden $E_{d}$ global symmetry~\cite{Cremmer:1979up, deWit:1985iy, deWit:1986mz, Berman:2013eva, Coimbra:2011ky}. 
These exceptional field theories are  based on extended spaces with natural $E_{d}\times R^+$ actions. As we have discussed above, the bosonic degrees of freedom can be parametrized by generalised metric $M_{MN}$, and the bosonic sector of M-theory action can be written in term of generalised metric as well.
The local $O(d)$ symmetry in double field theory is generalised to $H_d$, which is the maximally compact subgroup of $E_{d}$.

\section{$SL(5)$ Exceptional Field Theory}
\label{sec:sl5}

In $SL(5)$ exceptional field theory, we consider the eleven dimensional supergravity compactifies on an orientible four dimensional $\cal{M}$. The U-duality of M-theory is represented by $E_4 = SL(5)$ global symmetry in $SL(5)$ exceptional field theory~\cite{Hull:2007zu}. 
The corresponding maximal ungauged supergravity in seven dimensions is realized with global duality group $SL(5)$ as the hidden symmetry, with maximal subgroup $SO(5)$. The total fields are represented with the $7+4$ split given by~\cite{Musaev:2015ces} 
\begin{equation}
	\label{split}
	\left\{ g_{\m\n}, A_{\m\,a}, \phi_{ab}, C_{\m\n\r}, B_{\m\n\,a}, A_{\m\,ab}, \phi_{abc}  \right\},
\end{equation}
in which   $a, b$ run from 1 to 4 stands for the internal indices. 
The dynamics of the $14$ scalar fields $\phi_{ab}$ and $\phi_{abc}$ can be parametrised  in terms of matrix $V_i^{\a\b}$ in the coset $SL(5)/SO(5)$. 

In the corresponding generalised geometry construction, the generalised bundle for $SL(5)$ exceptional field theory is extended to $TM\oplus \Lambda^2~ T^*M$.
The section of the extended bundle is then a formal sum of $V\equiv \nu+\rho$ with a vector $\nu$ and a $2$-form $\rho$ which can be considered as an extended vector with $10$ components $V^I, I=1,...10$
\begin{equation}\label{vector}
	V^I=
	\left(\begin{matrix}
		\nu^i\\
		\rho_{ij}
	\end{matrix}\right)\, ,
\end{equation}
while $i,j =1,...,4$ and $\rho_{ij}=-\rho_{ji}$. 
The resulting $SL(5)$ exceptional field theory is defined on a $10$ dimensional extended space with coordinates $x^A$.  
The $10$ dimensional index $A$ can be considered as an antisymmetric pair of
indices in the fundamental representation of $SL(5)$ as 
$A\equiv [aa^{\prime}]$, $a,a^{\prime}=1,...,5$ \cite{Blair:2014zba}.

The $SL(5)$ bundle can be reduced to $SO(5)$ bundle by choosing the element of the coset $SL(5)/SO(5)$. This is equivalent to choosing a positive definite fibre metric $M_{AB}$ for each point $x\in \cal M$. 
The generalised metric is parametrised by a symmetric metric $g_{ij}$ transforming in the $\bf 10$ of $SO(4)\subset SO(5)$ and a $3$-form $C_{ijk}$ transforming as $\bf 4$ of $SO(4)$. 
The metric $g_{ij}$  and the $3$-form $C_{ijk}$ can be packaged into the so-called generalised metric  $M_{AB}$ as in double field theory~\cite{Duff:1990hn, Hull:2007zu}. This generalised metric serves as the overall metric for the extended space with coordinates $x^{A}$ where $A$ form the $10$ dimensional antisymmetric representation of $SL(5)$~\cite{Berman:2010is}. The generalised metric  $M_{AB}$~\eqref{genm} can be decomposed in terms of ``little generalised metric'' $m_{ab}$ with~\cite{Berman:2011cg, Park:2013gaj, Blair:2014zba}
\begin{equation}
M_{AB} \equiv M_{aa^\prime, bb^\prime} = m_{ab} m_{a^\prime b^\prime} - m_{ab^\prime} m_{a^ \prime b} \,,
\end{equation}
where $m_{ab}$ is symmetric tensor under $SL(5)$ transformation with  $m_{ij} = V_i^{\a\b} V_{j\,\a\b}$, which is also called as little generalized metric~\cite{Musaev:2015ces}.
This decomposition reduces the complexity of the parametrisation, while containing the exact degrees of freedom for the coset  $R^{+}\times SL(5)/SO(5)$. The  $R^{+}$ factor corresponds to extra scalar degree of freedom related to the wrapping factor of the reduction from eleven dimensional supergravity.

The action of $SL(5)$ exceptional field theory can be completely expressed with the little metric $m_{ab}$ as shown in~\cite{Berman:2010is, Park:2013gaj, Blair:2014zba}
\begin{equation}
	\begin{split}
		S &= \int_\Sigma |m|^{-1} \left( - \frac{1}{8} m^{ab} m^{a^\prime b^\prime}
		\partial_{a a^\prime} m^{c d} \partial_{b b^\prime} m_{c d} + \frac{1}{2} m^{ab}
		m^{a^\prime b^\prime} \partial_{aa^\prime} m^{cd} \partial_{cb^\prime} m_{bd}
		\right.\\
		& \quad \left. +\frac{1}{2} \partial_{aa^\prime} m^{ab} \partial_{bb^\prime}
		m^{a^\prime
			b^\prime} + \frac{3}{8}  m^{ab}  m^{a^\prime b^\prime}
		\partial_{aa^\prime}\ln |m| \, \partial_{bb^\prime} \ln |m| - 2 m^{a^\prime
			b^\prime} \partial_{aa^\prime} m^{ab} \partial_{bb^\prime} \ln|m| \right.
		\\
		& \quad \left. + m^{a^\prime
			b^\prime}\partial_{aa^\prime}\partial_{bb^\prime} m^{ab} - m^{ab}
		m^{a^\prime b^\prime} \partial_{a a^\prime} \partial_{bb^\prime} \ln |m| \right)
		\,, \label{actionsl5}
	\end{split}
\end{equation}
where  the determinant of the little metric is denoted by $m \equiv \det m_{ab}$, while $|m|^{-1}$ is $SL(5)$ singlet integral measure.
The little metric $m_{ab}$ is parametrized in terms of the usual metric $g_{ij}$ and the $3$-form $C_{ijk}$. 
The non-geometric background fields are represented by an alternative metric and a dual covector $\Omega^{ijk}$. This provides the convenience to study the duality transformation between the geometric and non-geometric frames in $SL(5)$ exceptional field theory. In particular, this little metric takes the form of
\eq{
\label{Gmetric}
{m}_{ab} = e^{-\phi/2}
 \begin{pmatrix}
  |g|^{-1/2}(g_{ij}+W_i V_j+V_i W_j+W_i W_j(1+V^2)) & V_i+W_i(1+V^2)  \\[0.1cm]
  V_j+W_j(1+V^2)  & |g|^{1/2}(1+V^2)
 \end{pmatrix}.
}
The scalar $\phi$ comes from the dimension reduction with $e^\phi=|g_7|^{1/7}$, where $g_7$ is the determinant of metric in the external seven dimensions. The vector $V^i$ is a dualisation of the $3$-form reads
\eq{\label{V}
	V^i=\frac{1}{3!}\varepsilon^{ijkl}C_{jkl},
}
and the covector $W_i$ is the dualisation of dual tri-vector of the $3$-form $\Omega^{ijk}$
\eq{\label{W}
	W_i=\frac{1}{3!}\varepsilon_{ijkl}\Omega^{jkl}.
}
The corresponding $SL(5)$ generalised vielbein is
\eq{
\label{}
{E}^\alpha{}_{a} =e^{-\phi/4}
 \begin{pmatrix}
  e^{-1/2}(e^\mu{}_i+V^\mu W_i) & e^{1/2}V^\mu  \\[0.1cm]
  e^{-1/2}W_i  & e^{-1/2}
 \end{pmatrix},
}
with its inverse 
\eq{
\label{}
{E}_\alpha{}^{a} =e^{\phi/4}
 \begin{pmatrix}
  e^{1/2}e_\mu{}^i & -e^{-1/2}W_\mu  \\[0.1cm]
  -e^{1/2}V^i  & e^{-1/2}(1+V^jW_j)
 \end{pmatrix}.
} And the generalised little metric~\eqref{Gmetric} can be expressed in terms of generalised vielbein $m_{ab}=E_a{}^\alpha\,S_\alpha{}^\beta\, E^\beta{}_{b}$.
By considering only the geometric usual metric $g_{ij}$ and the $3$-form $C_{ijk}$, or non-geometric background fields $\Omega^{ijk}$, one can reduce the little metric in geometric and non-geometric frames, respectively. 
In the geometric frame, $W_i=0$, the generalised little metric~\eqref{Gmetric} becomes
\eq{
	\label{Geometric}
	{m}_{ab} = e^{-\phi/2}
	\begin{pmatrix}
		|g|^{-1/2}g_{ij} & V_i \\[0.1cm]
		V_j & |g|^{1/2}(1+V^2)
	\end{pmatrix}.
}
In the so-called non-geometric frame which is the U-duality frame of the geometric frame, the generalised metric with $V^i=0$ instead,
reduces to
\eq{
	\label{NGmetric}
	{\tilde m}_{ab} = e^{-\phi/2}
	\begin{pmatrix}
		|\tilde g|^{-1/2}(\tilde g_{ij}+W_i W_j) & W_i \\[0.1cm]
		W_j  & |\tilde g|^{1/2}
	\end{pmatrix},
} where $e^\phi=|g_7|^{1/7}$ with $g_7$ is the determinant of external seven dimensional metric.
By applying $SL(5)$ transformation to the geometric little metric~\eqref{Geometric}, we study the U-duality transformation from geometric to non-geometric frame.

\subsection{$SL(5)$ Field Transformation}

In $SL(5)$ exceptional field theory, the U-duality transformation is represented via a $SL(5)$ group transformation. A $SL(5)$ transformation matrix $U$ acts on the generalised metric in the geometric frame $m_{ab}$ \eqref{Geometric} via conjugation, \emph{i.e.}, 
\eq{
\tilde{m}\, (\hat{g}, \hat{V})= U^t~ m(g, V)~ U
}
and therefore defines a field redefinition
\eq{
(g, V) \to (\hat{g}, \hat{V}).
}
We denote the general $SL(5)$ transformation matrix in the form of
\eq{\label{sl5m}
	{U}^{a}{}_b =
	\begin{pmatrix}
		a^i{}_j & b^i \\[0.1cm]
		c_j  & d
	\end{pmatrix}.
}
The little generalised metric in the non-geometric frame is realized via 
${\tilde m}_{ab}={U}_{a}{}^c\,m_{cd}\,{U}^d{}_{b}$ with the restriction that $ad-bc=1$. 
With ${U}_{a}{}^b$ as in the form of~\eqref{sl5m}, we obtain the upper-left component component of 	 ${\tilde m}_{ab}$ under $SL(5)$ transformation 
\eq{\label{ginter}
	|\tilde g|^{-1/2} ~\tilde g= |g|^{-1/2} a^t~ g~ a + c^t~( g~V)^{t}~a + a^t~(g~V)c  + |g|^{1/2}~ (1+V^2)~ c^t~ c ,
}
in which $V_i= g_{ij} V^j$ is denoted by $g V$ in the matrix form.
Furthermore, we have
\eq{\label{ginter2}
	\tilde g= |\tilde g|^{1/2} ~[|g|^{-1/2} a^t~ g~ a + c^t~( g~V)^{t}~a + a^t~(g~V)c  + |g|^{1/2}~ (1+V^2)~ c^t~ c].
}
This can be expressed in general mapping
\eq{
	\tilde g= |\tilde g|^{1/2} ~|g|^{-1/2}  \big[(a^t + |g|^{1/2} ~c^t~V^t  )~g~ ( a+ |g|^{1/2}~V~c )+  c^t~|g|~c \big].
}
By denoting $\gamma$ as
\eq{\label{rho}
	\gamma= a + |g|^{1/2} ~V~ c\,,
}
 we have $\tilde g_{ij}$ transformed to
\eq{
	\tilde g=|\tilde g|^{1/2} ~|g|^{-1/2}  \big[\gamma^t~ g~\gamma + c^t~|g|~ c \big],
}
and further manifest to~\footnote{Note that the density invariance under local Riemannian geometry $SO(5)$ requires that the determinant of the generalised metric to be invariant~\cite{Coimbra:2011ky}, we have
		$e^{-5\phi/2}~ |g|^{-1/2} = e^{-5\tilde\phi/2}~ \tilde{|g|}^{-1/2}$.
Recall the wrap factor can be defined according to the external metric $g_{IJ}$ of the transverse seven dimensions with $e^\phi\equiv|g_7|^{1/7}$, the prefactor  $|\tilde g|^{1/2} ~|g|^{-1/2}$ can be further simplified~\cite{Blair:2014zba}.}
\eq{\label{tildeg}
\tilde g=|\tilde g|^{1/2} ~|g|^{-1/2}  \gamma^t~ [g+(c~\gamma^{-1})^t~|g|~(c~\gamma^{-1})]~\gamma .
}
Note that there is an extra term $(c~\gamma^{-1})^t~|g|~(c~\gamma^{-1})$ added to the metric $g$ in 
 this U-dual mapping incorporated the dualisation of the $3$-form, vector $V^i$. 
Recall that  the mapping $\gamma^{-1}: T^*M \to TM \Leftrightarrow \gamma^{-1} \in TM\otimes TM$, 
the second term of $\tilde g$ on the right side overall gives $(c~\gamma^{-1})^t~|g|~(c~\gamma^{-1}) \in T^*M \otimes T^* M$, same as $g \in T^*M \otimes T^* M$ of the first term. 
In the U-dual field redefinition, the metric $g$ on $TM$ over the manifold $M$ transforms to the  metric on $E = TM \oplus \Lambda^2 T^*M$, while  the pullback  of the metric in $TM$ with extra contribution from  $\Lambda^2 T^*M$, such as $(c~\gamma^{-1})^t~|g|~(c~\gamma^{-1}) \in T^*M \otimes T^* M$ in $SL(5)$ exceptional field theory.
Thus, we have the metric under U-dual field transformation obtains extra contribution from membrane through $\Lambda^2 T^*M$  which is different with T-dual transformation in double field theory. Part of the work was started in~\cite{Sun:2016vnl}.

By reading off the upper-right element $e^{-\phi}  g_{ij} V^j $ of $m_{ab}$ under $SL(5)$ transformation , we have the vector(dualisation of 3-form $C_{jkl}$) transformed to
\eq{
	\tilde g~ \tilde V = |g|^{-1/2} a^t~ g~ b + c^t~( g~V)^{t}~b + a^t~(g~V)~d  + |g|^{1/2}~ (1+V^2)~ c^t~ d,
} and further reshuffled to 
\eq{
	\tilde{g}~\tilde V=  ~|g|^{-1/2}  \big[(a^t + |g|^{1/2} ~c^t~V^t  )~g~ ( b+ |g|^{1/2}~V~d )+  c^t~|g|~d \big].
} Therefore, the vector $V$ after transformation can be defined with
\eq{	\tilde V= \tilde{g}^{-1}  ~|g|^{-1/2}~ \gamma^t \big[~g~ ( b+ |g|^{1/2}~V~d )+  (\gamma^{t})^{-1}~c^t~|g|~d \big],}
Denoting the term on the right hand side in the bracket as $\sigma = g~ ( b+ |g|^{1/2}~V~d )+  (\gamma^{t})^{-1}~c^t~|g|~d$,  therefore, the dual vector 
\eq{\label{tildeV}
 \tilde{V} = 	\tilde{g}^{-1} |g|^{-1/2}~\gamma^t~ \sigma.
}
Note that here $\tilde{V}$ is still the dual 1-vector of 3-form $C_{jkl}$. To transform it into the dual 1-form of 3-vector $\Omega^{ijk}$ in the non-geometric dual frame, we define 
\eq{
	W = \tilde{g}~ \tilde{V}= |g|^{-1/2}~\gamma^t~ \sigma.
}
Therefore, the $SL(5)$ U-dual exceptional field theory\,(in terms of $\tilde{g}, W, \Omega$) is realized by the $SL(5)$ field transformations with general anchor in the form of $\rho=\gamma$.

\subsection{Examples of Non-geometric U-dual Frames}

Now we take particular $SL(5)$ transformations of the the generalized little matrix~\eqref{sl5m} and derive the resulting field redefinitions to the non-geometric U-dual frames.
Note that the off-diagonal components $b_i, c^j$  in the transformation matrix~\eqref{abm} involve the field transformations in between  $TM$ and $\Lambda^2 T^*M$ 
while $\Lambda^2 T^*M$ corresponding to M2-brane charge with $3$-form $C_{\mu \alpha_\beta}$  as the gauge fields. With non-trivial off-diagonal components of transformation matrix presented, $b$ in the transformation matrix denotes the non-trivial mapping $b: \Lambda ^2 T^* M \to TM \Leftrightarrow b \in \Lambda ^2 T M \otimes TM$. 

\subsubsection*{Frame I}

For the first example, we set the off-diagonal elements $b_i$ and $c^j$ in the general transformation matrix~\eqref{sl5m} to be trivial\,(introduce as \emph{weak constraint}), therefore the global transformation does not involve the transformations in between $\Lambda^2 T^*M$ and $TM$. The corresponding Courant algebroid anchor acts on the extended bundle $E=TM \oplus \Lambda^2 T^*M$ and map it  to the dual bundle $E^*=T^*M \oplus \Lambda^2 TM$ in the global manner.
This is analogy with applying Lie algebroid anchor mapping in double field theory, one obtains the redefined fields from $E=TM \to E=T^*M$~\cite{Blumenhagen:2013aia}. Here in $SL(5)$ exceptional field theory, as the membrane dependence incorporated, we have the Courant algebroid anchor mapping from $E=TM \oplus \Lambda^2 T^*M$ to $E^*=T^*M \oplus \Lambda^2 TM$~\cite{Courant:90, Gualtieri:2003dx, Berman:2011cg}.  

More precisely, we chooses the general transformation matrix ${U}$ in the form of
\eq{\label{sl5mad}
	{U} =
	\begin{pmatrix}
		a &~ 0 \\[0.1cm]
		0  &~ d
	\end{pmatrix}.
}
It acts on the little generalised metric as 
${\tilde m}={U^t} \,m\,{U}$. 
Therefore, we have the $SL(5)$ U-dual metric $\tilde g$ in \eqref{tildeg} transformed to 
\eq{
	\tilde g= |\tilde g|^{1/2} ~|g|^{-1/2}  ~a^t~ g~a,
}
and the $SL(5)$ U-dual form $W$ in \eqref{tildeV} reads
\eq{
	W = \tilde{g}~ \tilde{V}= a^t~g~V.
}
In particular, considering the index structure of $a^i{}_j$ and $d$, we take a natural choice of the transformation matrix~\eqref{sl5mad} 
\eq{\label{transl5}
	{U}^{a}{}_b =
	\begin{pmatrix}
	\delta^{i}{}_{j} & 0 \\[0.1cm]
	0	& \mathds 1
	\end{pmatrix},
}
and therefore we have the field redefinitions as\,(via $e^{\phi} \equiv |g_7|^{1/7}$)
\eq{
& \tilde{g}_{ij}=(1+V)^2\, g_{ij}, \\
& \tilde W_i = g_{ij} V^j .
}
Recall the dualization~\eqref{V},~\eqref{W}, and $\varepsilon_{ijkl}= |g|^{-1/2} \eta_{ijkl}$ , the dual tri-vector of the $3$-form $C_{mnp}$ therefore becomes
\eq{
\Omega^{ijk} = (1+V)^{-1} g^{im} g^{jn} g^{jp} C_{mnp},
} which is exactly the field redefinition of the $3$-form $C_{mnp}$ employed in~\cite{Blair:2014zba}.
The field strength of the 3-form $C_{ijk}$ can be defined as 
\eq{F_{ijkl}=4\,\partial_{[i}\,C_{jkl]},
} while the dual field strength of the 3-vector $\Omega^{ijk}$ is a totally non-geometric flux in the non-geometric U-dual frame. One can introduce a winding derivative 
\eq{
\tilde{\partial}^{ij} \equiv \frac{1}{2} \epsilon^{ijkl} \partial_{kl} = \partial^{ij}+\Omega^{ijk}\partial_k
}
on the dual field $\Omega^{ijk}$, once can define the non-geometric U-dual R-flux as
\eq{R^{i,jklm}=4 \,\tilde{\partial}^{i[j}\,\Omega^{klm]}.
}
More details of the M-theory fluxes under the $SL(5)$ transformation has been studied in~\cite{Blair:2014zba} with fluxes under U-duality chain given.
Based on the field redefinitions, by redefining the $SL(5)$ metric and 3-form $C_ijk$ with the dual fields, a non-geometric U-dual action can also be described with the non-geometric $\tilde{g}_{ij}$,  the U-dual tri-vector $\Omega^{ijk}$.

In this frame, we take~\eqref{transl5} as an example of transformation matrix with trivial off-diagonal elements, while different diagonal elements can be chosen, and result in different field transformations. 
Based on different field redefinitions, different non-geometric U-dual action can be derived.

\subsubsection*{Frame II}
For the second example, we consider a $SL(5)$ transformation characterized by a combination of a $V$- and $W$-transformation with the transformation matrix
\eq{\label{vw}
	{U}_{\text{V-W}} =
	\begin{pmatrix}
		\mathds{1} &~ V \\[0.1cm]
		-W	&~ 0
	\end{pmatrix},
}
for which in order to satisfy the $SL(5)$ properties, we have to require $W=V^{-1}$. The generalized metric under $V$- and $W$-transformation results from geometric little generalized metric~\eqref{Geometric} is 
\eq{
	\label{VW}
	{m}_{ab} = e^{-\phi/2}
	\begin{pmatrix}
		|g|^{-1/2}g_{ij}-W_i V_j-V_i W_j+W_i W_j(1+V^2) \ \ \ & 	|g|^{-1/2}V_i-W_i V^2  \\[0.1cm]
	|g|^{-1/2}	V_j- W_j V^2  & V^2
	\end{pmatrix}.
}
The non-geometric U-dual frame is represented by the metric $g$ and field $W_i$ instead as
\eq{
	\label{NGeo}
	\tilde{m}_{ab} = e^{-\phi/2}
	\begin{pmatrix}
		|g|^{-1/2}g -W g W^{-1}-W^{-1} g V+  |g|^{1/2} W W(1+V^2)\ \ \  & 	|g|^{-1/2} g W^{-1}-W W^2 \\[0.1cm]
		|g|^{-1/2} g W^{-1}-W W^2 & W^2
	\end{pmatrix},
} with the requirement $W=V^{-1}$.
Utilizing the transformation matrix~\eqref{vw} and the anchor mapping with~\eqref{tildeg} and~\eqref{tildeV},
we have 
\eq{
&\gamma =1-|g|^{1/2}\,V\,W,\\[0.1cm]
& |\tilde{g}|^{-1/2}\tilde{g} = 	|g|^{-1/2}g -W g W^{-1}-W^{-1} g V+  |g|^{1/2} W W(1+V^2)
}
and thus we have in the non-geometric frame the generalized metric is represented by
\eq{
	\label{NGeometric}
	\tilde{m}_{ab} = e^{-\phi/2}
	\begin{pmatrix}
		|\tilde{g}|^{-1/2}\tilde{g}_{ij} & W_i \\[0.1cm]
		W_j & |\tilde{g}|^{1/2}(1+W^2)
	\end{pmatrix}.
} 
Here the non-geometric U-dual frame results from the  $V$- and $W$-transformation is in the same form of the geometric generalised metric. The action in the non-geometric frame, with the U-dual metric on the cotangent bundle $T^*M$, and the U-dual of 3-form $\Omega^{ijk}$ specified with dual bundle $\Lambda^2 TM$, the action in the non-geometric U-dual frame can be derived.
Considering that the little metric transformed into the U-dual little metric $\tilde m$ in the non-geometric U-dual frame with the dual extended bundle $E*=T^*M \oplus \Lambda^2 TM$.

In the non-geometric U-dual frame, the action of $SL(5)$ exceptional field theory is thus expressed in terms of $\tilde{m}$ manifested in the form of~\eqref{actionsl5} under the global $SL(5)$ group transformation.

\subsubsection*{Frame III}

Regards to the U-duality chain in M-theory, in~\cite{Blair:2014zba} the $SL(5)$ element in question is given by
\eq{\label{udual}
	{U}^{a}{}_b =
	\begin{pmatrix}
		\delta^{i}{}_{j}-n^i \overline{n}_j & n^i \\[0.1cm]
		-\overline{n}_j	& 0
	\end{pmatrix},
} with vector $n^i$ specifies the directions in which the U-duality acts, where $n^i \overline{n}_i=1$.
The U-dual generalised metric under the U-duality chain is $m_{mn}={U}_{m}{}^a m_{ab}{U}^{b}{}_n$ with duality under the direction $a$ and $b$ to the U-dual direction $m$ and $n$.
By applying the anchor mapping we derived in~\eqref{tildeg} and~\eqref{tildeV}, the metric and dual fields under the U-duality version of Buscher rules can be directly derived. In heterotic double field theory, the anchor mapping also not only incorporates the T-duality via algebroid anchor mapping, but directly includes the first order $\alpha^\prime$ correction in the heterotic Buscher transformation~\cite{Blumenhagen:2014iua} in the convenient algebroid anchor mapping approach. By applying the $SL(5)$ Courant algebroid anchor mapping with~\eqref{udual}, the Buscher transformation can be extended to the U-duality Buscher transformation. The resulting U-dual fields, under each direction of U-duality, can be directly given by $SL(5)$ Courant algebroid anchor mapping of the fields with anchor $\rho=\gamma$~\eqref{rho}. The U-dual field strength $H_{ijkl}$ under each direction can be obtained via each application of transformation with transformation matrix~\eqref{udual} as well. 

By specifying the U-duality direction one by one with the T-duality directions, this anchor mapping can be reduced to the T-duality Buscher transformations on M-theory fluxes as shown in~\cite{Blair:2014zba}.

\section{$SO(5,5)$ Exceptional Field Theory}
\label{sec:so55}

The $SO(5,5)$ exceptional field theory is realized by compactifying 
the eleven dimensional supergravity on an orientible five dimensional manifold $\cal M$. The global symmetry group is $E_5=SO(5,5)$ representing the U-duality transformation. In addition to the $2$-form introduced in $SL(5)$ theory, an additional $5$-form is also introduced to the fibres, due to the appearance of M2-brane and M5-brane. Compared to the $SL(5)$ generalised  bundle $E=TM \oplus \Lambda^2 T^*M $, the generalised bundle is further extended to~ $E=TM \oplus \Lambda^2 T^*M \oplus \Lambda^5  T^*M$~\cite{Hull:2007zu, Berman:2011pe} and the so-called non-geometric dual bundle becomes $E^*=T^*M \oplus \Lambda^2 TM \oplus \Lambda^5 TM$.
A section of the bundle $E$ is then a formal sum of 
$U\equiv \nu+\rho+\sigma$ with a vector $\nu$, a $2$-form $\rho$ and a $5$-form $\sigma$. 
With the global duality group $SO(5,5)$, the ungauged maximal supergravity is realized in six dimensions~\footnote{Note that in the reduced supergravity action the $6$-forms do not appear, and thus we do not need to consider the Courant bracket of $5$-form $\sigma$ for studying the Courant algebroid structure in $SO(5,5)$ exceptional field theory in this section.}. The field content with the $6+5$ decomposition can be given by~\cite{Abzalov:2015ega}
\eq{
\{g_{\mu \nu}, A_{\mu\,m}, \phi_{mn}, C_{\mu \nu \rho}, B_{\mu \nu\,m} , A_{\mu\,mn}, \phi_{mnp}
\},
}
where $m,n$ are the internal indices running from $1$ to $5$. 
The dualisation of $3$-form $C_{\mu\nu \rho}$ can be defined as 
\eq{
	X_{\mu} = C_{\mu\nu \rho}V^{\nu \rho},
	\ \ \ \ \ \ \ \ \ \ 
	V^{\nu \rho} = \frac{1}{3!}\epsilon^{\nu \rho
		\alpha\beta\gamma}C_{\alpha\beta\gamma},
} in which $X_{\mu}$ is one $1$-form  in six dimensions, and
$\epsilon^{\rho\sigma\alpha\beta\gamma}$ is a totally antisymmetric tensor.
Combining the dual $1$-form $X_{\mu}$ with the five dual $1$-form $A_{\mu\,m}$ and ten $1$-form $A_{\mu\,mn}$, we obtain in total sixteen $1$-form fields, being the Majorana-Weyl spinor representation of $SO(5,5)$, $A_\mu^M, M=1,...16$. The bilinear form $B_{\mu\nu m}$ is then in the $\bf 5$ representation of $GL(5)\subset SO(5,5)$. Moreover, the $25$ scalar fields $\phi_{mn}$ and $\phi_{mnp}$ are unified into a $16$ by $16$ matrix $V_M^{\alpha \alpha^\prime}$ parametrising the coset $SO(5,5)/(SO(5)\times SO(5))$, with $\alpha, \alpha^\prime=1,...4$ as the spinor indices for each $SO(5)$ and $M,N=1,...16$ as the $SO(5,5)$ spinor indices for the extended space. While $V_M^{\alpha \alpha^\prime}$ acts as the generalised vielbein,  the generalised metric ${\cal M}_{MN}$ on the extended space is defined as~\cite{Abzalov:2015ega}
\eq{
{\cal M}_{MN}=V_M^{\alpha \alpha^\prime}\, V_{N\,\alpha \alpha^\prime},
}
with the inverse of the generalised vielbein defined by
\eq{
V_M^{\alpha \alpha^\prime}\,V_{\alpha \alpha^\prime}^N=\delta_M^N,\ \ \ \
 V_M^{\alpha \alpha^\prime}\,V_{\beta \beta^\prime}^M=\delta{^\alpha_\beta}\,\delta^{\alpha^\prime}_{\beta^\prime}.
}
The generalised metric $\mathcal{M}_{MN}$ can be parametrised 
in terms of the metric $g_{\mu \nu}$ and the $3$-form potential of $SO(5,5)$ exceptional field theory $C_{\mu\nu\rho}$ by~\cite{Berman:2011pe, Thompson:2011uw, Musaev:2013kpa, Abzalov:2015ega}
{\small
\eq{\label{SO55m}
	\mathcal{M}_{MN} = 
	\begin{pmatrix}
		g_{\mu \nu} + \frac{1}{2}C_{\mu}^{\ \rho \sigma}C_{\nu \rho \sigma} + 
		\frac{1}{16}X_{\mu}X_{\nu} & \ \ \
		\frac{1}{\sqrt{2}}C_{\mu}^{\ \nu_{1} \nu_{2}} + \frac{1}{4\sqrt{2}}X_{\mu}V^{\nu_{1}\nu_{2}} & 
		\frac{1}{4}|g|^{-\frac{1}{2}}X_{\mu} \\[0.1cm]
		\frac{1}{\sqrt{2}}C^{\mu_{1}\mu_{2}}_{\ \ \ \ \ \nu} + \frac{1}{4\sqrt{2}}V^{\mu_{1}\mu_{2}}X_{\nu} & \ \ \ 
		g^{\mu_{1}\mu_{2},\nu_{1}\nu_{2}} + 
		\frac{1}{2}V^{\mu_{1}\mu_{2}}V^{\nu_{1}\nu_{2}} & \ \ \
		\frac{1}{\sqrt{2}}|g|^{-\frac{1}{2}}V^{\mu_{1}\mu_{2}} \\[0.1cm] 
		\frac{1}{4}\,|g|^{-\frac{1}{2}}X_{\nu} & 
		\frac{1}{\sqrt{2}}\,|g|^{-\frac{1}{2}}V^{\nu_{1}\nu_{2}} & 
		|g|^{-1} 
	\end{pmatrix}, 
}
}
in which $g^{\mu_{1}\mu_{2},\nu_{1}\nu_{2}}=\frac{1}{2}(g^{\mu_{1}\nu_{1}} g^{\mu_{2}\nu_{2}}-g^{\mu_{1}\nu_{2}} g^{\mu_{2}\nu_{1}})$ contracts with antisymmetric pair of indices by raising the indices.
The coordinates are parametrised by sixteen internal coordinates with spinorial indices $\mathbb{X}^{M},\, M=1,...,16$.
Utilizing the branching rules for \textbf{16}\,, we have the  decomposition of  $\mathbb{X}^{M}$ as:
\be
\mathbb{X}^{M} \ \longrightarrow\  \{\ 
x^{\mu}\,,y^{[2]}_{\mu\nu}\,,z^{[5]}_{\mu\nu\rho\sigma\tau}\ \}\,,
\ee
where $x^{\mu}$ denotes the usual geometric coordinate, and $y_{\mu\nu}$, $z_{\m\n\rho\sigma\tau} $ correspond to the winding modes of the M2- and M5-branes, respectively. 
Under section condition utilized in~\cite{Berman:2011pe},
 all fields will be independent of the additional coordinates $y^{[2]}_{\mu\nu}\,,z^{[5]}_{\mu\nu\rho\sigma\tau}$, that 
\eq{
\frac{\partial}{\partial_{y_{ab}}}=0\,,\ \ \
\frac{\partial}{\partial_{z_{\mu\nu\rho\sigma\tau}}}=0. 
}
Here, instead of this section condition, we will later discuss the \emph{weak constraint} we introduced in $SL(5)$ theory. By setting the off-diagonal elements in the transformation matrix to be trivial, one can produce the action on the bundle $E=TM \oplus \Lambda^2 T^*M \oplus \Lambda^5  T^*M$ with transformation restricted in between $\Lambda^n TM$ and $\Lambda^n T^*M$. The so-called non-geometric U-dual theory is then defined on $E^*=T^*M \oplus \Lambda^2 TM \oplus \Lambda^5 TM$. Namely, the M-theory reduced $SO(5,5)$ theory mapped to another U-dual M-theory with dual M2- and M5-brane dependences incorporated.

\subsection{$SO(5,5)$ Field Transformation}
The Lagrangian of $SO(5,5)$ exceptional field theory is parametrised by the generalised metric $M_{MN}$ in terms of the usual metric and $3$-form given in~\cite{Berman:2011pe}
{\small 
\begin{multline}\label{LagSO5}
	L = g^{1/2} \Biggl(\frac{1}{16} M^{MN} (\partial_M M^{KL})( \partial_N
	M_{KL} ) - {\frac{1}{2}} M^{MN} (\partial_N M^{KL}) (\partial_L M_{MK}) \\
	+\frac{3}{128}   M^{MN} (M^{KL} \partial_M M_{KL})(M^{RS} \partial_N M_{RS}) 
	-\frac{1}{8}    M^{MN} M^{PQ}(M^{RS} \partial_P M_{RS}) (\partial_M M_{NQ}) 
	\Biggr), 
\end{multline}
} 
where $\partial_M = \bigl(\frac{\partial}{\partial x^a},
\frac{\partial}{\partial y_{ab}}, \frac{\partial}{\partial z} \bigr)$ containing the dependences of M2- and M5-branes. 
Up to integration by parts, the Lagrangian in terms of the generalised metric~\eqref{LagSO5} incorporates the diffeomorphism and gauge transformation leads to the supergravity Lagrangian as~\eqref{lag}.  
To map the $SO(5,5)$ exceptional theory in a global manner to its U-dual theory, we choose the general $SO(5,5)$ transformation matrix of the form
\eq{
U_{M}{}^{N} = 
\begin{pmatrix}\label{abm}
a_{\m}{}^{\n} & b_{\m\n_{1}\n_{2}} & m_{\m} \\[0.1cm]
k^{\m_{1}\m_{2}\n} & d^{\m_{1}\m_{2}}_{\n_{1}\n_{2}} & n^{\m_{1}\m_{2}} 
\\[0.1cm]
p^{\n} & q_{\n_{1}\n_{2}} & z
\end{pmatrix}\,,
}
with restriction that  $det\, U=1$. It acts on the generalised metric~\eqref{SO55m} via conjugation in the form of
\eq{
\tilde{ M}\, (\tilde{g}, \tilde{C}, \tilde{X}, \tilde{V})= U^t~ { M}(g, C, X, V)~ U
}
and therefore defines a field redefinition 
\be
(g, C, X, V) \to (\tilde{g}, \tilde{C},\tilde{X},\tilde{V}),
\ee
resulting in the U-dual $SO(5,5)$ exceptional field theory on  the bundle $E^*=T^*M \oplus \Lambda^2 TM \oplus \Lambda^5  TM$. 

For convenience of deriving the anchor mapping, we choose to apply
 the $SO(5,5)$ transformation to the inverse of $SO(5,5)$ generalised metric 
{\small
\eq{\label{SO55m-inverse}
	{M}^{MN} = 
	\begin{pmatrix}
		g^{\mu \nu}  &
		-\frac{1}{\sqrt{2}}C^{\mu}_{\ \nu_{1} \nu_{2}} & 
		\frac{\sqrt{g}}{4} X^{\mu} \\[0.1cm]
		-\frac{1}{\sqrt{2}}C_{\mu_{1}\mu_{2}}^{\ \  \ \ \nu} & 
	\ \ \	g_{\mu_{1}\mu_{2},\nu_{1}\nu_{2}} + 
		\frac{1}{2}C_{\mu_{1}\mu_{2}}^{\ \ \ \ \ \mu}\  C_{\nu_{1}\nu_{2}\mu} & 
	\ \ \	-\frac{\sqrt{g}}{\sqrt{2}}V_{\mu_{1}\mu_{2}}-\frac{\sqrt{g}}{4\sqrt{2}}C_{\mu_{1}\mu_{2}\mu} \,X^\mu \\[0.1cm] 	\frac{\sqrt{g}}{4} X^{\nu} & 
	\ \ \	-\frac{\sqrt{g}}{\sqrt{2}}V_{\nu_{1}\nu_{2}}-\frac{\sqrt{g}}{4\sqrt{2}}C_{\nu_{1}\nu_{2}\mu} \,X^\mu & 
	\ \  \ 	1+ \frac{1}{2}V_{\mu\nu}\,V^{\mu\nu}+\frac{1}{16}\,X^{\mu}\,X_{\mu}
	\end{pmatrix}, 
}
}
in the way of
\be
{{M}^{MN}}^{\prime} = U^{M}{}_{K}{ M}^{KL}U_{L}{}^{N} \,.
\ee
The upper left component is the inverse of standard metric $g^{\mu \nu}$ mapped to $\tilde{g}^{\mu \nu}$, and in matrix form $\tilde{g}^{\mu \nu}$ takes the form of
{\small
	\eq{
		\tilde{g}^{-1}=&  \left(-\frac{a^t C~ g^{-1}}{\sqrt{2}\, } +k^t \left(\frac{C^2~g^{-1}}{2}+ (g\otimes g)_A \right)+p^t \left(-\frac{C \sqrt{|g|}\, X~g^{-1}}{4 \sqrt{2}}-\frac{\sqrt{|g|} \,g^2 V}{\sqrt{2}}\right)\right) k \\[0.1cm]
		&+ \left(\frac{a^t \sqrt{|g|}\, X ~g^{-1}}{4}+k^t \left(-\frac{C \sqrt{|g|}\, X~g^{-1}}{4 \sqrt{2}}-\frac{\sqrt{|g|}\, g^2 V}{\sqrt{2}}\right)+p^t \left(\frac{g^2 V^2}{2}+\frac{X^2 g^{-1}}{16}+1\right)\right) p\\[0.1cm]
		& + \left(a^t~g^{-1}-\frac{k^t C ~g^{-1}}{\sqrt{2}}+\frac{p^t\sqrt{|g|} X g^{-1} }{4}\right) a.
	}
}
Here $g$ represents the matrix form of $g_{\mu\nu}$, while $g^{-1}$ represents the matrix form of $g^{\mu\nu}$.
The inverse of the U-dual metric $\tilde{g}^{-1}$ can be expressed with $SO(5,5)$ anchor and an extra $\delta$ term as
\eq{\label{invg}
	\tilde{g}^{-1}= \gamma^t\, g^{-1}\, \gamma\,+ \delta_1\,,
}
which can be reshuffled again to 
\eq{\label{invg2}
	\tilde{g}^{-1}= \gamma^t\,[\, g^{-1}\,+ (\gamma^t)^{-1}\, \delta_1 \,\gamma^{-1}\, ] \, \gamma,
} with $SO(5,5)$ Courant algebroid anchor 
\eq{
\gamma =a-\frac{k\, C}{\sqrt{2}}+\frac{1}{4} \,p \,\sqrt{|g|} \, X,
}
and 
\eq{
\delta_1=&k^{t} (g\otimes g)_A \,k +p^{t}(\frac{V^2\,g^2}{2}+\frac{X^2 \,g^{-1}}{16} -\frac{|g|\, X^2 \,g^{-1}}{16} +1)\,p
\\[0.2cm]
&-\frac{k^t\, \sqrt{|g|}\, V\, g^2 \, p }{\sqrt{2}}-\frac{p^t \,\sqrt{|g|}\, V \,g^2 k  }{\sqrt{2}}\,,
}
in which $(g\otimes g)_A
=g_{\mu_{1}\mu_{2},\nu_{1}\nu_{2}}=\frac{1}{2}(g_{\mu_{1}\nu_{1}} g_{\mu_{2}\nu_{2}}-g_{\mu_{1}\nu_{2}} g_{\mu_{2}\nu_{1}})$ lowering an antisymmetric pair of indices.
The inverse mapping for the metric $g^{\mu\nu}$ is simply defined by the inverse of  $\gamma$ as 
\eq{
	{g}^{-1}= (\gamma^t)^{-1}\, (\tilde{g}^{-1}\, - \delta_1\,) \gamma^{-1}\,.
}
Here the non-trivial $\delta_1$ comes from the non-trivial off-diagonal components $b_i, c^j$  in the transformation matrix~\eqref{abm}. The field redefinitions with  $\delta_1$ involve the transformation between $\Lambda^2 T^*M$,  $\Lambda^5 T^*M$
and $TM$. And in the inverse field redefinition from U-dual frame to geometric frame, this involve the transformation from $\Lambda^2 TM$,  $\Lambda^5 TM$
to $TM$ instead.
With the elements $k^{\m_{1}\m_{2}\n}, p^{\n}, b_{\m\n_{1}\n_{2}}, m_{\m}$, in the $SO(5,5)$ transformation matrix  $U_{M}{}^{N}$ set to be trivial, the transformation in between $\Lambda^2 ~T^*M \oplus \Lambda^5 ~ T^*M$ and $TM$ are not involved. This can be manifested via $\delta_1 = 0$, and while the higher forms are not involved,  the $SO(5,5)$ field transformation can be reduced to the same form of type II and heterotic Lie algebroid anchor mapping shown in~\cite{Blumenhagen:2013aia, Blumenhagen:2014iua}.

For the field redefinition of $3$-form $C_{\mu\nu\rho}$, we read off the upper middle component of the transformed generalised metric, and have 
{\small 
\eq{
	-\frac{1}{\sqrt{2}}\, \tilde C \, \tilde g^{-1}=&  \left(-\frac{a^t C g^{-1} }{\sqrt{2}}+k^t \left(\frac{C^2 g^{-1}}{2 }+ (g\otimes g)_A \right)+p^t \left(-\frac{C \sqrt{|g|}\, X g^{-1}}{4 \sqrt{2} }-\frac{\sqrt{|g|} \,g^2 V}{\sqrt{2}}\right)\right) d \\[0.1cm]
	&+ \left(\frac{a^t \sqrt{|g|}\, X g^{-1} }{4}+k^t \left(-\frac{C \sqrt{|g|}\, X g^{-1}}{4 \sqrt{2}}-\frac{\sqrt{|g|}\, g^2 V}{\sqrt{2}}\right)+p^t \left(\frac{g^2 V^2}{2}+\frac{X^2 g^{-1}}{16}+1\right)\right) q\\[0.1cm]
	& + \left(a^t g^{-1}-\frac{k^t C g^{-1}}{\sqrt{2}}+\frac{p^t\sqrt{|g|} X g^{-1}}{4}\right) b,
}
}
which can be reshuffled to
\eq{
	-\frac{1}{\sqrt{2}}\, \tilde C \, \tilde g^{-1}&= 
	\gamma^t\, g^{-1}\, \sigma\,+ \delta_2\,,	
}
 by defining
\eq{
	\delta_2=
	&k^{t} (g\otimes g)_A \,d +p^{t}(\frac{V^2\,g^2}{2}+\frac{X^2 \,g^{-1}}{16} -\frac{|g|\, X^2 \,g^{-1}}{16} +1)\,q
	\\[0.2cm]
	&-\frac{k^t\, \sqrt{|g|}\, V\, g^2 \, q }{\sqrt{2}}-\frac{p^t \,\sqrt{|g|}\, V \,g^2 d  }{\sqrt{2}}\,,
		\\[0.2cm]
	\sigma =& b-\frac{ C\, d}{\sqrt{2}}+\frac{1}{4}  \,\sqrt{|g|} \, X\, q,
}
Taking the $SO(5,5)$ U-duality transformation with 
$	\gamma =a-\frac{k\, C}{\sqrt{2}}+\frac{1}{4} \,p \,\sqrt{|g|} \, X$, we have the dual $3$-form $C_{\mu\nu\rho}$ transformed into
\eq{
	\tilde{C}= - \sqrt{2} \,\gamma^t\,[\, g^{-1}\,+ (\gamma^t)^{-1}\, \delta_2 \,\sigma^{-1}\, ] \, \sigma \,\tilde{g}.
}
Now it is convenient to define the dual antisymmetric $3$-vector $\Omega^{\sigma\tau\kappa}$ from the dual $3$-form $C_{\mu\nu\rho}$ as follows:
\eq{
	\tilde{\Omega}=\tilde{C}~ \tilde{g}^{-1}\tilde{g}^{-1}\tilde{g}^{-1}.
}
In order to determine the redefined 1-form $X_\mu$, it is convenient to check the transformed upper-right component $\frac{\sqrt{g}}{4} X^{\mu}$ in $M^{MN}$~\eqref{SO55m-inverse}. The $SO(5,5)$ U-duality transformation leads to
{\small 
\eq{
\frac{\sqrt{|\tilde{g}|}}{4} \tilde X\, \tilde g^{-1}=&  \left(-\frac{a^t C  g^{-1}}{\sqrt{2} }+k^t \left(\frac{C^2  g^{-1}}{2 }+ (g\otimes g)_A \right)+p^t \left(-\frac{C \sqrt{|g|}\, X  g^{-1}}{4 \sqrt{2} }-\frac{\sqrt{|g|} \,g^2 V}{\sqrt{2}}\right)\right) n \\[0.1cm]
	&+ \left(\frac{a^t \sqrt{|g|}\, X  g^{-1}}{4}+k^t \left(-\frac{C \sqrt{|g|}\, X  g^{-1}}{4 \sqrt{2}}-\frac{\sqrt{|g|}\, g^2 V}{\sqrt{2}}\right)+p^t \left(\frac{g^2 V^2}{2}+\frac{X^2  g^{-1}}{16}+1\right)\right) z\\[0.1cm]
	& + \left(a^t  g^{-1}-\frac{k^t C  g^{-1}}{\sqrt{2}}+\frac{p^t\sqrt{|g|} X  g^{-1}}{4} \right) m.
}
}
Recall the $SO(5,5)$ U-dual transformation with 
$	\gamma =a-\frac{k\, C}{\sqrt{2}}+\frac{1}{4} \,p \,\sqrt{|g|} \, X$, we have 
\eq{
	\frac{\sqrt{|\tilde{g}|}}{4} \tilde X\, \tilde g^{-1}&= 
	\gamma^t\, g^{-1}\, \eta\,+ \delta_3\,,
}
with non-trivial $\delta_3$ and $\eta$ term
\eq{
	\delta_3=
	&k^{t} (g\otimes g)_A \,n +p^{t}(\frac{V^2\,g^2}{2}+\frac{X^2 \,g^{-1}}{16} -\frac{|g|\, X^2 \,g^{-1}}{16} +1)\,z
	\\[0.2cm]
	&-\frac{k^t\, \sqrt{|g|}\, V\, g^2 \, z }{\sqrt{2}}-\frac{p^t \,\sqrt{|g|}\, V \,g^2 n  }{\sqrt{2}}\,,
	\\[0.2cm]
	\eta =& m-\frac{ C\, n}{\sqrt{2}}+\frac{1}{4}  \,\sqrt{|g|} \, X \, z.
}
This can be further reshuffled, and the U-dual of $1$-form $X_\mu$ in six dual dimensions transformed to
\eq{
	\tilde{X}= \frac{4}{\sqrt{|\tilde{g}|}} \,\gamma^t\,[\, g^{-1}\,+ (\gamma^t)^{-1}\, \delta_3 \,\eta^{-1}\, ] \, \eta \,\tilde{g}.
}
The dual $1$-vector $\cal{\tilde{X}}$ (as the dualisation of $3$-vector $\Omega^{\sigma\tau\kappa}$) can furthermore be defined as
\eq{
	{\cal \tilde{X}}= \tilde{X}\,\tilde{g}^{-1} ,
}
where
\eq{
		{\cal \tilde{X}}^{\mu} = \Omega^{\mu\nu \rho} \ 	{\cal \tilde{V}}_{\nu \rho},
	\ \ \ \ \ \ \ \ \ \ 
	{\cal \tilde{V}}_{\nu \rho} = \frac{1}{3!}\epsilon_{\nu \rho\alpha\beta\gamma}\Omega^{\alpha\beta\gamma}.
} 
in which ${\cal \tilde{X}}^{\mu}$ is one $1$-vector  in the U-dual framework, and the background for the $3$-vector $\Omega^{\alpha\beta\gamma}$ would be U-fold comparing to the T-fold background in double field theory.
Therefore, all the fields appearing in $SO(5,5)$ exceptional field theory are defined under the Courant algebroid anchor mapping and resulting in the U-dual background of $SO(5,5)$ exceptional field theory.

\subsection{Examples of Non-geometric U-dual Frames}

Now we take particular $SO(5,5)$ transformations of the the generalized matrix $M^{MN}$~\eqref{SO55m-inverse} and derive the resulting field redefinitions to the non-geometric U-dual frames.

For the simple example with \emph{weak constraint} as we introduced for $SL(5)$ exceptional field theory, we set the off-diagonal elements in the general transformation matrix~\eqref{abm} to be trivial, therefore the global transformation does not involve the transformations in between $\Lambda^m \,TM$ and $\Lambda^n\,T^*M$ with different $m$ and $n$. The corresponding Courant algebroid anchor acts on the extended bundle $E=TM \oplus \Lambda^2 T^*M \oplus \Lambda^5 T^*M$ and map it  to the dual bundle $E^*=T^*M \oplus \Lambda^2 TM \oplus \Lambda^5 TM$ in the global manner.
The fields under the Courant algebroid anchor mapping, with  $\delta_1 = \delta_2 =\delta_3 = 0$,  are redefined by 
\eq{
	& \tilde{g}=  a^{-1}~g~(a^t)^{-1}\,,
	\\[0.2cm]
	&	\tilde{\Omega}= 
	\,a^t \, g^{-1} \,{d\, C} \, \tilde{g}^{-1}\,\tilde{g}^{-1} ,
	\\[0.2cm]
	& {\cal \tilde{X}}= 
	\frac{\sqrt{|g|}}{\sqrt{|\tilde{g}|}}  \,\,a^t \, g^{-1} \, z  \, X  .
}

In particular, considering the index structure of $a_{\m}{}^{\n}$, $d^{\m_{1}\m_{2}}_{\n_{1}\n_{2}}$ and $z$ we take a simple choice of the transformation matrix~\eqref{abm} 
\eq{
U_{M}{}^{N} =
	\begin{pmatrix}
		\delta_{\m}{}^{\n}\ \ \ & 0 &\ \ 0 \\[0.1cm]
		0 \ \ \ &  \epsilon^{\m_{1}\m_{2}}_{\n_{1}\n_{2}} & \ \  0 
		\\[0.1cm]
		0 \ \ \ & 0 & \ \  \mathds 1
	\end{pmatrix}\,,
} and we can define the dual $3$-vector $\tilde{\Omega}^{\sigma\tau\kappa}$ and $1$-vector ${\cal \tilde{X}}^\mu$ as
\eq{
	&	\tilde{\Omega}^{\sigma\tau\kappa}=  \tilde{g}^{\sigma\mu} \tilde{g}^{\tau\nu} \tilde{g}^{\kappa\rho} \tilde{C}_{\mu\nu\rho},
	\\[0.2cm]
	& {\cal \tilde{X}}^\mu= \tilde{g}^{\mu\nu} X_\nu .
}
The Lagrangian of $SO(5,5)$ theory~\eqref{LagSO5} can then be mapped to the U-dual ``non-geometric'' theory in the extended space under the global $SO(5,5)$ group transformation. The U-dual fields $\tilde{g^{ij}}, \Omega^{ijk}, \tilde{X}^i$ lives in the bundle $E^*=T^*M \oplus \Lambda^2 TM \oplus \Lambda^5 TM$. Instead of  winding modes of the M2- and M5-branes, momentum modes appear with coordinates $y^{\mu\nu}$, and $z^{\m\n\rho\sigma\tau}$.
Therefore, the dual generalised metric in the dual extend bundle $E^*$, under the global $SO(5,5)$ global transformation, is parametrized by the  U-dual fields $\tilde{g^{ij}}, \Omega^{ijk},  {\cal \tilde{X}}^i$ as
{\small
	\eq{\label{USO55m}
		\mathcal{M}^{MN} = 
		\begin{pmatrix}
			\tilde g_{\mu \nu} + \frac{1}{2} \Omega^{\mu \rho \sigma}\Omega^{\nu}{}_{\rho \sigma} + 
			\frac{1}{16}{\cal \tilde{X}}^{\mu}{\cal \tilde{X}}^{\nu} & \ \ \
			\frac{1}{\sqrt{2}}\Omega^{\mu}_{\ \nu_{1} \nu_{2}} + \frac{1}{4\sqrt{2}}{\cal \tilde{X}}^{\mu}{\cal \tilde{V}}_{\nu_{1}\nu_{2}} & 
			\frac{1}{4}|\tilde g|^{-\frac{1}{2}}{\cal \tilde{X}}^{\mu} \\[0.1cm]
			\frac{1}{\sqrt{2}}\Omega_{\mu_{1}\mu_{2}}^{\ \ \ \ \ \nu} + \frac{1}{4\sqrt{2}}{\cal \tilde{V}}_{\mu_{1}\mu_{2}}{\cal \tilde{X}}^{\nu} & \ \ \ 
			\tilde g_{\mu_{1}\mu_{2},\nu_{1}\nu_{2}} + 
			\frac{1}{2}{\cal \tilde{V}}_{\mu_{1}\mu_{2}}{\cal \tilde{V}}_{\nu_{1}\nu_{2}} & \ \ \
			\frac{1}{\sqrt{2}}|\tilde g|_{-\frac{1}{2}}{\cal \tilde{V}}_{\mu_{1}\mu_{2}} \\[0.1cm] 
			\frac{1}{4}\,|\tilde g|^{-\frac{1}{2}}{\cal \tilde{X}}^{\nu} & 
			\frac{1}{\sqrt{2}}\,|\tilde g|^{-\frac{1}{2}}{\cal \tilde{V}}_{\nu_{1}\nu_{2}} & 
			|\tilde g|^{-1} 
		\end{pmatrix}, 
	}
}
 with
${\cal \tilde{X}}^{\mu} = \Omega^{\mu\nu \rho} \ 	{\cal \tilde{V}}_{\nu \rho},
{\cal \tilde{V}}_{\nu \rho} = \frac{1}{3!}\epsilon_{\nu \rho\alpha\beta\gamma}\Omega^{\alpha\beta\gamma}.
$

The Lagrangian of $SO(5,5)$ exceptional field theory~\eqref{LagSO5} in the U-dual frame can be expressed with the U-dual fields parametrized into U-dual generalised metric $M_{MN}$ in~\eqref{LagSO5}. 
Here the underline structure of the U-duality transformation is governed by the Courant algebroid as the anchor acts on the extended bundle $E=TM \oplus \Lambda^2 T^*M \oplus \Lambda^5  T^*M$. 
By taking the general form of the U-dual transformation with non-trivial $\delta_1, \delta_2, \delta_3$, we not just consider the field transformation in between $\Lambda^n TM$ and $\Lambda^n T^*M$, yet consider the full U-dual field transformations in between the extended generalised bundle $E=TM \oplus \Lambda^2 T^*M \oplus \Lambda^5  T^*M$ and $E^*=T^*M \oplus \Lambda^2 TM \oplus \Lambda^5  TM$. 

By choosing trivial off-diagonal components of transformation matrix as we have in this example, it is obvious that  the dual metric in the non-geometric U-dual frame $\tilde{g}=  a^{-1}~g~(a^t)^{-1}$ only involve the transformation $TM \to T^*M$ without the extra $\delta$-term contribution from $\Lambda^2 T^*M$ or $\Lambda^5 T^*M$. If one considers the mapping $TM \to T^*M$, $\Lambda^2 T^*M \to \Lambda^2 TM$, individually, the Courant algebroid reduces to a structure as Lie algebroid in double field theory, and Lie bi-algebroid~\cite{Blumenhagen:2013aia, Hitchin:2003cxu, Gualtieri:2003dx}.
However, considering the extended U-duality mapping from  $E=TM \oplus \Lambda^2 T^*M \oplus \Lambda^5  T^*M$ to $E^*=T^*M \oplus \Lambda^2 TM \oplus \Lambda^5  TM$, we need to utilize Courant algebroid to provide the overall anchor mapping. Different with $SL(5)$ exceptional field theory, in $SO(5,5)$ theory, there is the extra M5-brane embedded from M-theory, and one may consider whether the Courant algebroid is still closed with Jacobi identity satisfied. In Section~\ref{sec:algebroid}, we have shown that with the gauge 3-form presented, the Courant algebroid is closed with Jacobi identity satisfied since the Jacobiator takes an exact form. Although in $SO(5,5)$ theory, a 6-form gauge field exists as the electromagnetic dual of the $3$-form, $d\tilde{C}_6 \sim * d C_3$~\cite{Cremmer:1998px}, it does not contribute to the kinetic term of the $SO(5,5)$ Lagrangian explicitly. Only the 3-form appears in the kinetic term of M-theory~\eqref{kin} and the generalised metric~\eqref{SO55m}. Therefore, the Courant algebroid structure we discussed in Section~\ref{sec:algebroid} still holds in $SO(5,5)$ exceptional field theory.

\section{Higher Dimensional Exceptional Field Theories}
\label{sec:e67}

We have  shown that explicit anchor mapping, representing U-duality, can be found in $SL(5)$ and $SO(5,5)$ exceptional field theories. 
In this section, we study whether the U-duality anchor mapping still exists in the $E_{6}$ and $E_{7}$ exceptional field theories.

\subsection{E6 Exceptional Field Theory}

When we consider the eleven dimensional supergravity reduction on a $d$ dimensional manifold $\cal M$ for $d=6, 7$, the U-duality group is represented by the exceptional group  $E_{6}$ and $E_{7}$, respectively~\cite{Hull:2007zu, Berman:2011pe, Hohm:2013vpa,Hohm:2013uia}. Compared to the standard generalised geometry with generalised bundle $E=TM \oplus T^*M$, the generalised bundle is further extended to $E=TM \oplus \Lambda^2 T^*M \oplus \Lambda^5 T^*M\oplus \Lambda^6  TM$ and the so-called non-geometric U-dual bundle will be $E^*=T^*M \oplus \Lambda^2 TM \oplus \Lambda^5  TM \oplus \Lambda^6  T^*M$.
The coset  $E_6/H_6$ and  $E_7/H_7$ can be parametrised by a metric $g_{ij}$, $3$-form $C_{ijk}$, and a $6$-form $C_{i_1, ... i_6}$, and thus the generalised metric $\cal H$ and vielbein $E_{M}{}^{N}$ can be parametrised with the above fields as well. 

In particular, for the reduction to five dimensional supergravity, the U-duality group is represented by the $E_{6}$ exceptional group. The generalised vielbein is parametrised into~\cite{Berman:2011jh, Musaev:2013kpa}
\be\label{bein6}
E_{M}{}^{N} = 
({\rm det}e)^{-1/2}
\begin{pmatrix}
	e_{\mu}{}^{i} & - {\frac{1}{\sqrt{2}}} e_{\mu}{}^{j} C_{j i_{1} i_{2}} &
	{\frac{1}{2}} e_{\mu}{}^{i_{3}} U + {\frac{1}{4}} e_{\nu}{}^{i_{3}} C_
	{\mu j k} V^{\nu j k} \\
	0 & e^{\mu_1}{}_{[i_{1}} e^{\mu_{2}}{}_{i_{2}]} & -{\frac{1}{\sqrt{2}}} e^
	{\mu_{1}}{}_{j_{1}} e^{\mu_{2}}{}_{j_{2}} V^{j_{1} j_{2} i_{3}} \\
	0 & 0 & ({\rm det}e)^{-1} e_{\mu_{3}}^{i_{3}}
\end{pmatrix}\,,
\ee
and the generalised metric can be given by ${M}_{MN} = E_{M}{}^{K}\delta_{KL}E^{L}{}_{N}$, written in terms of the metric $g_{ij}$, $3$-form $C_{ijk}$, fields $V^{ijk}$, and $U$, where the fields $V^{ijk}$ and $U$ are the dualisation of  $3$-form $C_{ijk}$, and a $6$-form $C_{i_1, ... i_6}$ explicitly with 
\be
\begin{split}
	U=\frac{1}{6}\epsilon^{ijklmn}C_{ijklmn},\\
	V^{ikl}=\frac{1}{3!}\epsilon^{iklmnj}C_{mnj}.
\end{split}
\ee
The dynamics term $\sqrt{g} \left(R -{{1}\over{48}} {F^{(4)}}^{2}\right)$ can be parametrised in terms of this generalised metric up to integration by parts
\begin{multline}
	{{1}\over{24}} \, M^{MN} (\partial_M M^{KL})( \partial_N
	M_{KL} ) -{{1}\over{2}} \, M^{MN} (\partial_N M^{KL}) (\partial_L M_
	{MK}) \\
	+ {{19}\over{9720}} \, M^{MN} (M^{KL} \partial_M M_{KL})(M^{RS}
	\partial_N M_{RS}).
	\label{(E6lag)}
\end{multline}
This $E_6$ exceptional field theory Lagrangian is parametrised by generalised metric given by ${M}_{MN} = E_{M}{}^{K}\delta_{KL}E^{L}{}_{N}$, and thus parametrised by the metric $g_{ij}$, $3$-form $C_{ijk}$, fields $V^{ijk}$, and scalar $U$.  The fields $V^{ijk}$ and $U$ are the dualisation of  $3$-form $C_{ijk}$, and $6$-form $C_{i_1, ... i_6}$ which contract with the totally antisymmetric tensor $\epsilon^{ijklmn}$. However, the 6-form potential is not dynamical in $E_6$ exceptional field theory. Although the 6-form presents as the hodge dual $U$ in the generalised metric, by verifying the kinetic term of $E_6$ Lagrangian~\eqref{(E6lag)}, the overall gauge invariant field strength contribution vanishes.
Therefore, for the Courant Jacobiator evaluated for $E_6$ exceptional field theory, only the $3$-form  $C_{ijk}$ changing under gauge transformation matters. The Courant algebroid structure, which we have discussed in Section~\ref{sec:algebroid}, is sufficient for the existence of Courant algebroid in  $E_6$ exceptional field theory. 

\subsection{E7 Exceptional Field Theory}

For the reduction to four dimensional supergravity, the U-duality group is represented by $E_7$ group. The generalised vielbein $E_{M}{}^{N}$ is given
by~\cite{Berman:2011jh, Musaev:2013kpa, Hohm:2013uia}
{\small
	\begin{equation}\label{bein7}
		e^{-{{1}\over{2}}}
		\begin{pmatrix}
			e_{\mu}{}^{i} & - {{1}\over{\sqrt{2}}} e_{\mu}{}^{j} C_{j i_1 i_2} &
			{{1}\over{\sqrt{2}}} e_{\mu}{}^{[i_3} U^{i_4]} + {{1}\over{4\sqrt
					{2}}}  e_{\mu}{}^{j} X_{j}{}^{i_3 i_4} & {{1}\over{2}} e_{\mu}{}^{j}
			C_{j i_5 k }  U^{k} - {{1}\over{24}} e_{\mu}{}^{j} X_{j}{}^{kl} C_
			{kl i_5}\\
			0 & e^{\mu_1}{}_{[i_1} e^{\mu_{2}}{}_{i_2]} & -{{1}\over{\sqrt{2}}} e^
			{\mu_1}{}_{j_1} e^{\mu_{2}}{}_{j_2} V^{j_1 j_2 i_3 i_4} & {{1}\over
				{\sqrt{2}}} e^{\mu_1}{}_{[j} e^{\mu_{2}}{}_{i_5]} U^{j} + {{1}\over{4
					\sqrt{2}}} e^{\mu_1}{}_{j_1} e^{\mu_{2}}{}_{j_2} X_{i_5 }{}^{j_1
				j_2} \\
			0 & 0 & \textup{e}^{-1} e_{\mu_3}{}^{[i_3} e_{\mu_{4}}{}^{i_4]} & -
			{{1}\over{\sqrt{2}}} e_{\mu_3}{}^{j_1} e_{\mu_{4}}{}^{j_2} C_{j_1 j_2
				i_5} \\
			0 & 0 & 0 & \textup{e}^{-1} e^{\mu_5}{}_{i_5} 
		\end{pmatrix},
\end{equation}} where $e$ represents the determinant of the vielbein, $ g_{\mu \nu} = e^{i}_{\mu} e^{j}_{\nu} \eta_{ij}$, $V^{i_1 \dots i_4}$ represents the dualisation of the $3$-form potentials, $V^{i_1 \dots i_4}= {{1}\over{3!}} \epsilon^{i_1 \dots i_4, j_1 \dots  j_3} C_{j_1 \dots j_3}$
 and $U^{i}$ is the dualisation of the $6$-form potentials 
$U^{i}= {{1}\over{6!}} \epsilon^{i j_1 \dots j_6} C_{j_1 \dots j_6}$, with $X_{i}{}^{jk}= C_{ilm} V^{jklm}$, in which the Greek indices are tangent space indices, and the Latin indices are space indices. 

The governing Lagrangian of $E_7$ exceptional field theory is again parametrised by generalised metric given by ${M}_{MN} = E_{M}{}^{K}\delta_{KL}E^{L}{}_{N}$, and thus parametrized by the metric $g_{ij}$, $3$-form $C_{ijk}$ as
\begin{multline}
	{{1}\over{48}} \, M^{MN} (\partial_M M^{KL})( \partial_N
	M_{KL} ) - {{1}\over{2}} \, M^{MN} (\partial_N M^{KL}) (\partial_L M_
	{MK}) \\
	+ {{17}\over{37632}} \, M^{MN} (M^{KL} \partial_M M_{KL})(M^{RS}
	\partial_N M_{RS}),
\end{multline}
where the dualisation fields  $U^{i}$ , $V^{i_1 \dots i_4}$  contract with the seven dimensional totally antisymmetric tensor $\epsilon_{i_1\dots i_7}$. This reproduces the dynamics term 
\eq{
\sqrt{g} \left( R - {{1}\over{48}} {F^{(4)}}^{2} - {{1}\over{8!}}
{F^{(7)}}^{2} \right),
}
in which $F^{(4)}$ is the field strength of the 3-form as for the lower dimensional exceptional field theory,
 and $F_{(7)}$ is the field strength of the 6-form defined as 
\eq{ F^{(7)}_{\mu_1 \dots \mu_7} = 7 \partial_{[\mu_1} C_{\mu_2 \dots
	\mu_7]} + 140 C_{[\mu_1 \dots \mu_3} \partial_{\mu_4} C_{\mu_5 \dots
	\mu_7]}.
} 
By such field  parametrization, the Lagrangian is manifested with a $6$-form gauge field as the electromagnetic dual of the $3$-form with $d\tilde{C}_6 \sim * d C_3$~\cite{Cremmer:1998px}. However, this duality between the 4-form and 7-form field strengths can only be realised in eleven dimensions, and yet the $E_7$ exceptional field theory is seven-dimensional. Therefore, the  electromagnetic duality $d\tilde{C}_6 \sim * d C_3$ comes across a defect. And due to the appearance of 7-form field strength arising from 6-form potential, the govern Courant algebroid also needs to be further checked with Courant bracket by  incorporating the 6-form in Courant algebra~\eqref{Courant} and Jacobi identity~\eqref{bid}. We hope to come back to the discussion of  algebroid structure in  $E_7$ exceptional field theory in a future work.

\subsection{$E_d$ Symmetry  in M-theory}

For M-theory generalisation of the extended generalised geometry, the corresponding extended tangent bundle is extended with $\Lambda^2 \,T^*M$ corresponding to a M2-brane charge, $\Lambda^5 \,T^*M$ corresponding to M5-brane\,(for n>4),   $\Lambda^6 \,TM$ corresponding to the Kaluza-Kein monopole charge(for $5<n\leq 7$)~\cite{Hull:2006qs}. The $E_8$ exceptional field theory was constructed in~\cite{Hohm:2014fxa}.
For $E_6$ and $E_7$ exceptional field theories, the extended tangent bundle is $E=TM \oplus \Lambda^2 \,T^*M \oplus \Lambda^5\, T^*M\oplus \Lambda^6\,TM$. The coset space $E_n/H_n$ are parametrised by the metric $g$, a $3$-form $C$ associated with the $3$-form gauge field of eleven dimensional supergravity, and a $6$-form gauge field as the electromagnetic dual of the $3$-form, $d\tilde{C}_6 \sim * d C_3$. The action $E_6$ and $E_7$  on these fields includes the transformation between $3$-form $C_3$ and  $6$-form $\tilde{C}_6$.
Meanwhile, the M5-brane charge is the electromagnetic dual of the M2-brane charge, which is consistent with that the $5$-form not appearing in $SO(5,5)$ exceptional field theory. 

Under the field redefinition governed by the  Courant algebroid anchor mapping, all the exceptional fields can be transformed to their U-dual fields for $SL(5)$, $SO(5,5)$ and $E_6$ exceptional field theory. Therefore, all the elements in the generalized vielbein $E_{M}{}^{N}$~\eqref{bein6},\eqref{bein7} of $E_6$ and $E_7$ exceptional field theories can be expressed in terms of the U-dual fields through Courant algebroid anchor mapping. 
These fields in the dual exceptional field theory will result in the structure of $\tilde{g}= \rho^t\, g\, \rho\,+ \delta$. 
These extra $\delta$ terms appear due to the transformation from $\Lambda^p T^*M$ to $TM$ with the contribution from the membrane incorporated as for $SL(5)$ and $SO(5,5)$ exceptional field theory. 

From M-theory  point of view, our study  indicates that a U-dual M${}^\prime$-theory can be constructed via a global U-duality. From the generalised geometry point of view, this corresponds to the $E_d$ mapping from the extended bundle $E=TM \oplus \Lambda^2 T^*M \oplus \Lambda^5  T^*M\oplus \Lambda^6 TM$ to the U-dual bundle  $E^*=T^*M \oplus \Lambda^2 TM \oplus \Lambda^5 TM \oplus \Lambda^6  T^*M$ for $d\leq 7$. Analogous to the construction in double field theory with generalized bundle $TM \oplus T^*M$, when we consider the generalized bundle $E \oplus E^*$, this indicates a doubled effective theory of M-theory  may be incorporated.

As another interesting aspect, by considering the $E_d$ global symmetry in exceptional field theory in $d$ dimension, the U-duality is manifested as $E_d$ group transformation. The resulting U-dual exceptional field theory is in $d$ dimension(in the non-geometric U-dual framework) as well. Recall that the $E_d$ exceptional field theory is realized by compactifing the eleven dimensional supergravity on $T^d$, the external dimension is $11-d$.
Namely,  the total dimension\,(incorporating the doubled exceptional field theory in $E \oplus E^*$ and the external dimension) will results in $15, 16, 17, 18, 19$ total dimension with U-dual $E_4, E_5, E_6, E_7, E_8$ exceptional field theories present.

\section{Conclusion and Outlook}

In this paper, we studied the U-duality transformation in exceptional field theories with Courant algebroid anchor mapping. 
In Section~\ref{sec:algebroid}, we have shown that the closure of Courant bracket, and trivial Courant Jacobi identity can be realized in exceptional field theory with 3-form potential present.
By considering the eleven dimensional supergravity compactified on an orientible $d$ dimensional manifold $\mathcal{M}$ the global symmetry becomes $E_d$ group, which represents U-duality in exceptional field theory.
Take $SL(5)$ and $SO(5,5)$ exceptional field theories as examples, we  explicitly derive the U-dual field redefinition with Courant algebroid anchor mapping given.  We show that these U-duality transformations do exist\,(which not only incorporates T-duality but also S-duality) via the dependence of membrane incorporated. 

In particular, with the Lie algebroid generalised to Courant algebroid, we show that one can incorporate the dependence of membrane and obtain a global anchor mapping with the  extended bundle $E=TM \oplus \Lambda^2 \,T^*M$ to its dual extended bundle $E^*=T^*M \oplus \Lambda^2 \,TM$. 
Instead of  choosing the strong version of section condition, such as $\frac{\partial}{\partial y_{ij}} =0 $, we chose  the general section condition for the closure of Courant bracket 
$$
	\partial_{[x} \partial_{y]} X =0 \qquad \text{and} \qquad  \partial_{[x} X \cdot \partial_{y]} Y  =0,
$$
in which $y$ denotes the dependences of the $p$ form in $TM\oplus \Lambda^p T^*M$. 
The closure of Courant bracket is realized with the above section condition, and the Courant Jacobi identity is realized while being evaluated on the $3$-form field $C_{ijk}$ which is the only field changes under a gauge transformation.

By introducing a \emph{weak constraint} with trivial off-diagonal element in the transformation matrix, the field redefinitions are restricted to $\Lambda^n TM$ and $\Lambda^n T^*M$ with the same $n$, while $n=1, 2, 5, 6$. We show that the Courant algebroid anchor mapping reduces to Lie algebroid anchor mapping as in double field theory with the sub-bundle $TM \to TM^*$ from $E=TM \oplus \Lambda^2 \,T^*M \oplus \Lambda^5\, T^*M$. 
With all the fields explicitly redefined by their U-dual fields, the Lagrangian of exceptional field theories can be expressed in terms of their U-dual fields in the $E^* = T^*M \oplus \Lambda^2 \,TM \oplus \Lambda^5\, TM$ extended bundle as well.

For $E_6$ and $E_7$ exceptional field theories, the corresponding extended geometry is $E=TM \oplus \Lambda^2 \,T^*M \oplus \Lambda^5\, T^*M\oplus \Lambda^6\,TM$. We found that for $E_6$ exceptional field theory, the fields change under gauge transformation is still the 3-form potential $C_3$ parametrized  in the Lagrangian. Thus the U-duality can also be represented by Courant algebroid anchor mapping. 
While the Lagrangian $E_7$ exceptional field theories  with the 6-form potential $C_6$ presented, a further instigation on the Courant algebroid anchor mapping is needed.

In summary, we show that in exceptional field theory the U-dual field transformation is governed by Courant algebroid with membrane dependence incorporated.
From M-theory point of view,  the corresponding exceptional field theory  transforms under global U-duality group $E_d$, and indicates  a U-dual M-theory may exist within a dual extended bundle $E^*$.

\begin{acknowledgments}

The author thanks Ralph Blumenhagen, Falk Hassler, Weikun He,
Chris Hull, Qiang Jia, Jeong-Hyuck Park, Sheng Meng, Yi-Nan Wang and Jie Zhou for useful discussions. This work is supported by KIAS Individual Grant PG080701 and PG080704.

\end{acknowledgments}

\bibliography{reference}


\end{document}